\documentclass[longauth,useAMS,usenatbib]{aa}  

\usepackage{graphicx}
\usepackage{txfonts}
\usepackage{hyperref}
\usepackage{longtable}
\usepackage{appendix}

\begin{document} 

   \title{Revisiting Proxima with ESPRESSO \thanks{Based [in part] on Guaranteed Time Observations collected at the European Southern Observatory under ESO programme 1102.C-0744. by the ESPRESSO Consortium.} \thanks{This work makes use of observations from the LCOGT network.} \thanks{The data used in this paper is available in electronic form at the CDS via anonymous ftp to cdsarc.u-strasbg.fr (130.79.128.5) or via http://cdsweb.u-strasbg.fr/cgi-bin/qcat?J/A+A/.}}

  \author{A. Su\'{a}rez Mascare\~{n}o\inst{1,7} \and
         J. P. Faria   \inst{2,14} \and 
         P. Figueira \inst{2,17} \and
         C. Lovis \inst{10} \and
         M. Damasso \inst{11} \and
         J. I. Gonz\'alez Hern\'andez \inst{1,7} \and
         R. Rebolo \inst{1,7,16} \and
         S. Cristiani \inst{9} \and
         F. Pepe \inst{10} \and
         N. C. Santos \inst{2,14} \and
         M. R. Zapatero Osorio \inst{12} \and
         V. Adibekyan \inst{2,14}\and
         S. Hojjatpanah \inst{2,14}\and
         A. Sozzetti \inst{11} \and
         F. Murgas \inst{1,7} \and
         M. Abreu \inst{3,20} \and
         M. Affolter \inst{4} \and
         Y. Alibert \inst{5} \and
         M. Aliverti \inst{6} \and
         R. Allart \inst{10} \and
         C. Allende Prieto \inst{1,7}\and
         D. Alves \inst{3,20} \and
         M. Amate \inst{1} \and
         G. Avila \inst{8} \and
         V. Baldini \inst{9} \and
         T. Bandi \inst{4} \and
         S. C. C. Barros \inst{2} \and
         A. Bianco \inst{6} \and
         W. Benz \inst{4} \and
         F. Bouchy \inst{10} \and
         C. Broeng \inst{4} \and
         A. Cabral \inst{3,20} \and
         G. Calderone \inst{9} \and
         R. Cirami \inst{9} \and
         J. Coelho \inst{3,20} \and
         P. Conconi \inst{6} \and
         I. Coretti \inst{9} \and
         C. Cumani \inst{8} \and
         G. Cupani \inst{9} \and
         V. D'Odorico \inst{9, 18} \and
         S. Deiries \inst{8} \and
         B. Delabre \inst{8} \and
         P. Di Marcantonio \inst{9} \and
         X. Dumusque \inst{10} \and
         D. Ehrenreich \inst{10} \and
         A. Fragoso \inst{1} \and
         L. Genolet \inst{10} \and
         M. Genoni \inst{6} \and
         R. G\'{e}nova Santos \inst{1,7} \and
         I. Hughes \inst{10} \and
         O. Iwert \inst{8} \and
         F. Kerber \inst{8} \and
         J. Knusdstrup \inst{8} \and
         M. Landoni \inst{6} \and
         B. Lavie \inst{10} \and
         J. Lillo-Box \inst{12} \and
         J. Lizon \inst{8} \and
         G. Lo Curto \inst{8} \and
         C. Maire \inst{10} \and
         A. Manescau \inst{8} \and
         C. J. A. P. Martins \inst{2,14} \and
         D. M\'{e}gevand \inst{10} \and
         A. Mehner \inst{8} \and
         G. Micela \inst{13} \and
         A. Modigliani \inst{8} \and
         P. Molaro \inst{9,15} \and 
         M. A. Monteiro \inst{2} \and
         M. J. P. F. G. Monteiro \inst{2,14} \and
         M. Moschetti \inst{6} \and
         E. Mueller \inst{8} \and
         N. J. Nunes \inst{3,20} \and
         L. Oggioni \inst{6} \and
         A. Oliveira \inst{3,20} \and
         E. Pallé \inst{1,7} \and
         G. Pariani \inst{3} \and
         L. Pasquini \inst{8} \and
         E. Poretti \inst{6,15} \and
         J. L. Rasilla \inst{1} \and
         E. Redaelli \inst{6} \and
         M. Riva \inst{6} \and
         S. Santana Tschudi \inst{1} \and
         P. Santin \inst{9} \and
         P. Santos \inst{3,20} \and
         A. Segovia \inst{10} \and
         D. Sosnowska \inst{10} \and
         S. Sousa \inst{2} \and
         P. Span\`{o} \inst{6} \and
         F. Tenegi \inst{1} \and
         S. Udry \inst{10} \and
         A. Zanutta \inst{6} \and
         F. Zerbi \inst{6}        }

   \institute{Instituto de Astrof\'{i}sica de Canarias, E-38205 La Laguna, Tenerife, Spain\\
              \email{asm@iac.es}            \and                      
               Instituto de Astrofísica e Ciências do Espaço, Universidade do Porto, CAUP, Rua das Estrelas, 4150-
762 Porto, Portugal \and 
Instituto de Astrofísica e Ciências do Espaço, Universidade de Lisboa, Edif\'{i}cio C8, 1749-016 Lisboa, Portugal \and
Physikalisches Institut, Universität Bern, Siedlerstrasse 5, 3012      Bern, Switzerland \and 
Physikalisches Institut \& Center for
Space and Habitability, Universität Bern, Gesellschaftsstrasse 6, 3012 Bern, Switzerland \and 
INAF -  Osservatorio Astronomico di Brera,      Via Bianchi 46,         23807,         Merate,         Italy \and 
Departamento de Astrof\`{i}sica, Universidad de La Laguna, E-38206 La
Laguna, Tenerife, Spain \and 
European Southern Observatory,  Karl-Schwarzschild-Strasse 2,   85748,  Garching b. München,    Germany \and 
INAF -  Osservatorio Astronomico di Trieste,    Via Tiepolo 11, 34143,  Trieste,        Italy \and 
Observatoire astronomique de l’Université de Genève, 51 chemin des Maillettes, 1290 Versoix, Switzerland \and 
INAF - Osservatorio Astrofisico di Torino, Via Osservatorio 20, I-10025 Pino Torinese, Italy \and 
Centro de Astrobiología (CSIC-INTA), Carretera de Ajalvir km 4,  E-28850 Torrejón de Ardoz, Madrid, Spain \and 
INAF - Osservatorio Astronomico di Palermo, Piazza del
Parlamento 1, 90134 Palermo, Italy \and 
Departamento de Física e Astronomia, Faculdade de Ciências, Universidade do Porto, Rua do Campo
Alegre, 4169-007 Porto, Portugal. 7 \and 
Institute for Fundamental Physics of the Universe, Via Beirut 2, 34151 Miramare, Trieste,
Italy \and 
Consejo Superior de Investigaciones Científicas, 28006 Madrid, Spain \and 
European Southern Observatory, Alonso de Córdova 3107, Vitacura, Región Metropolitana, Chile \and 
Scuola Normale Superiore, Piazza dei Vavalieri 7, I-56126, Pisa, Italy \and 
Centro de Astrof\'{\i}sica da Universidade do Porto, Rua das Estrelas, 4150-762 Porto, Portugal \and
Faculdade de Ciências da Universidade de Lisboa (Departamento de Física), Edifício C8, 1749-016 Lisboa
}

   \date{A\&A Accepted - May 2020}

 
  \abstract
   {The discovery of Proxima b marked one of the most important milestones in exoplanetary science in recent years. Yet the limited precision of the available radial velocity data and the difficulty in modelling the stellar activity calls for a confirmation of the Earth-mass planet.} 
   {We aim to confirm the presence of Proxima b using independent measurements obtained with the new ESPRESSO spectrograph, and refine the planetary parameters taking advantage of its improved precision.}
   {We analysed 63 spectroscopic ESPRESSO observations of Proxima (Gl 551) taken during 2019. We obtained radial velocity measurements with a typical radial velocity photon noise of 26 cm$\cdot$s$^{-1}$. We combined these data with archival spectroscopic observations and newly obtained photometric measurements to model the stellar activity signals and disentangle them from planetary signals in the radial velocity (RV) data. We ran a joint Markov chain Monte Carlo analysis on the time series of the RV and full width half maximum of the cross-correlation function to model the planetary and stellar signals present in the data, applying Gaussian process regression to deal with the stellar activity signals.}
   {We confirm the presence of Proxima b independently in the ESPRESSO data and in the combined ESPRESSO+HARPS+UVES dataset. The ESPRESSO data on its own shows Proxima b at a period of 11.218 $\pm$ 0.029 days, with a minimum mass of   1.29 $\pm$ 0.13 M$_{\oplus}$. In the combined dataset we measure a period of 11.18427 $\pm$ 0.00070 days with a minimum mass of 1.173 $\pm$ 0.086 M$_{\oplus}$. We get a clear measurement of the stellar rotation period (87 $\pm$ 12 d) and its induced RV signal, but no evidence of stellar activity as a potential cause for the 11.2 days signal. We find some evidence for the presence of a second short-period signal, at 5.15 days with a semi-amplitude of only 40 cm$\cdot$s$^{-1}$. If caused by a planetary companion, it would correspond to a minimum mass of 0.29 $\pm$ 0.08 M$_{\oplus}$.  We find that for the case of Proxima, the full width half maximum of the cross-correlation function can be used as a proxy for the brightness changes and that its gradient with time can be used to successfully detrend the RV data from part of the influence of stellar activity.  The activity-induced RV signal in the ESPRESSO data shows a trend in amplitude towards redder wavelengths. Velocities measured using the red end of the spectrograph are less affected by activity, suggesting that the stellar activity is spot dominated. This could be used to create differential RVs that are activity dominated and can be used to disentangle activity-induced and planetary-induced signals. The data collected excludes the presence of extra companions with masses above 0.6 M$_{\oplus}$ at periods shorter than 50 days.}
  {}


\maketitle 
%

\section{Introduction}

   The discovery by \citet{AngladaEscude2016} of a planetary candidate orbiting the habitable zone of our closest neighbour, Proxima Centauri (Gl 551),  shook the planetary community as few other discoveries have done in recent years. It not only showed that the nearest star to the Sun could host a planetary system but also that, given the right conditions, it could host a habitable rocky planet \citep{Ribas2016}. This year the announcement of a second planetary candidate in the system brought attention back to Proxima \citep{Damasso2020}, making it even more interesting. 
   
   High precision radial velocity (RV) measurements give astronomers the possibility of detecting low-mass exoplanets, down to the mass of the Earth. Unfortunately, photospheric and chromospheric phenomena on the stellar surface, associated to the presence and evolution of magnetic fields, induce RV variations which, if stable over a few rotation periods, can mimic a planetary signal. Recognising and characterising them is key to disentangling true planet-induced signals \citep{QuelozHenry2001,  Dumusque2012, Santos2014, Robertson2014, Faria2019}. In the case of an M-dwarf, those variations would be of the order of a few m$\cdot$s$^{-1}$ even for quiet stars \citep{Masca2018}, larger than the amplitude measured for Proxima b. These activity-induced RV variations, combined with the limitations of previous generation spectrographs when studying mid-type M-dwarfs, and the importance of the discovery, make confirming Proxima b a task for the new ESPRESSO spectrograph \citep{Pepe2010}. The larger collecting power and improved precision helps to disentangle the different signals that coexist in the data, leading to a better characterisation of all the phenomena involved. 
  
 In this work we first analyse the newly obtained ESPRESSO data, searching for evidence of the presence of Proxima b. We  then combine the data with the archival data to perform a joint analysis combining activity proxies and RVs. Later we discuss the observed activity signals, their relationship with the RV measurements and the observed wavelength dependency of the velocities. We end with a discussion on the presence of additional short period planets in the system. 


\section{Proxima}

The closest star to the Sun is the smallest member of the Alpha Centauri system. It is an M5.5 star with a mass of just 0.12 M$_{\odot}$ \citep{Delfosse2000, Mann2015} and a radius of 0.15 R$_{\odot}$ \citep{Boyajian2012}. Its habitable zone ranges from distances of 0.05 AU to 0.1 AU \citep{AngladaEscude2016}. It is $\sim$1000 times less luminous than the Sun, which even at its close distance makes it invisible to the naked eye ($m_{V}$ $\sim$ 11.13). Proxima shows a very slow rotation of $\sim$ 83 days \citep{Benedict1998, Masca2016}, and a long-term activity cycle with a period ofapproximately seven years \citep{Masca2016, Wargelin2017}. Proxima is famous for its frequent and intense flares. It is estimated that it flares at least twice every day \citep{Davenport2016}. Those flares routinely increase the brightness of the star by 10-50\% (with one recorded case of a flare that made it visible to the naked eye) by increasing its flux by a factor of 40 \citep{HowardWard2018}. All these characteristics make Proxima a challenging but interesting target. Table~\ref{tab:parameters} shows the stellar parameters of Proxima.

\begin{table}
\begin{center}
\caption{Stellar properties of Proxima. \label{tab:parameters}}
\begin{tabular}[center]{l l l}
\hline
Parameter & Proxima & Ref. \\ \hline
RA (J2000) & 14:29:42.95 & 1 \\
DEC (J2000) & -62:40:46.17 & 1\\
$\mu \alpha$ ($mas$ yr$^{-1}$)& -3781.306 & 1 \\
$\mu \delta$ ($mas$ yr$^{-1}$)& 769.766 & 1 \\
Parallax ($mas$) &  768.50 $\pm$ 0.20 & 1\\
Distance (pc) & 1.3012 $\pm$ 0.0003 & 1 \\
$m_{B}$  (mag) & 12.95 $\pm$ 0.03 & 2 \\
$m_{V}$  (mag) & 11.13 $\pm$ 0.01 & 2 \\
$m_{Ks}$ (mag) & 4.384 $\pm$ 0.0033 & 9 \\
$M_{Ks}$ (mag) & 8.813 $\pm$ 0.0033 & 10 \\
Spectral Type  & M5.5V & 3\\
\textit{L}$_{*}$/\textit{L}$_{\odot}$ & 0.0016 $\pm$ 0.0006 & 4 \\
\textit{T}$_{\rm eff}$ (K) & 2900 $\pm$ 100 & 5 \\
$[Fe/H]$ (dex) & 0.0 $\pm$ 0.1 & 5 \\
M$_{*~D00}$ (M$_{\odot}$) & 0.120  $\pm$ 0.015 & 6 \\
M$_{*~M15}$ (M$_{\odot}$) & 0.1221   $\pm$ 0.0022 & 7 \\
\textit{R}$_{*}$ (\textit{R}$_{\odot}$) & 0.141  $\pm$ 0.021 & 4 \\
\textit{P}$_{\rm rot}$ (days) & $\sim$83.2 &8 \\
\textit{log g} (cgs) & 5.0 $\pm$ 0.25 & 5\\
\textit{log}$_{10}$ (\textit{R}$_{\rm HK}^{'})$ & -- 4.98 $\pm$ 0.13 & 0 \\
\hline
\end{tabular}
\end{center}
\textbf{References:} 0 - This work -- based on \citet{Masca2016}, 1 - \citet{Gaia2016}, 2 - \citet{WeiChun2014}, 3 - \citet{Bessell1991}, 4 -\citet{Boyajian2012}, 5 - \citet{Pavlenko2017}, 6 - Estimated using~ \citet{Delfosse2000}  (D00), 7 - Estimated using~ \citet{Mann2015} (M15), 8 - \citet{Masca2016}, 9 - \citet{Cutri2003}, 10 - \citet{Kervella2017}. \\
\textbf{Note:} Most works on Proxima b have relied on the stellar mass derived from \citet{Delfosse2000}.The measurement derived using \citet{Mann2015} is more precise, so we will use it as our primary measurement, but we also continue to calculate the planetary mass using \citet{Delfosse2000} to allow for a direct comparison with previous measurements.  
\end{table}

Proxima was selected to be part of the ESPRESSO Guaranteed Time Observations (GTO) survey that monitors a selected group of nearby stars ranging from early G-type to mid M-type \citep{Hojjatpanah2019}. The main goal of the survey is to discover low-mass planets in the habitable zone of their parent stars.

\section{Observations }

\begin{figure*}
        \includegraphics[width=18.0cm]{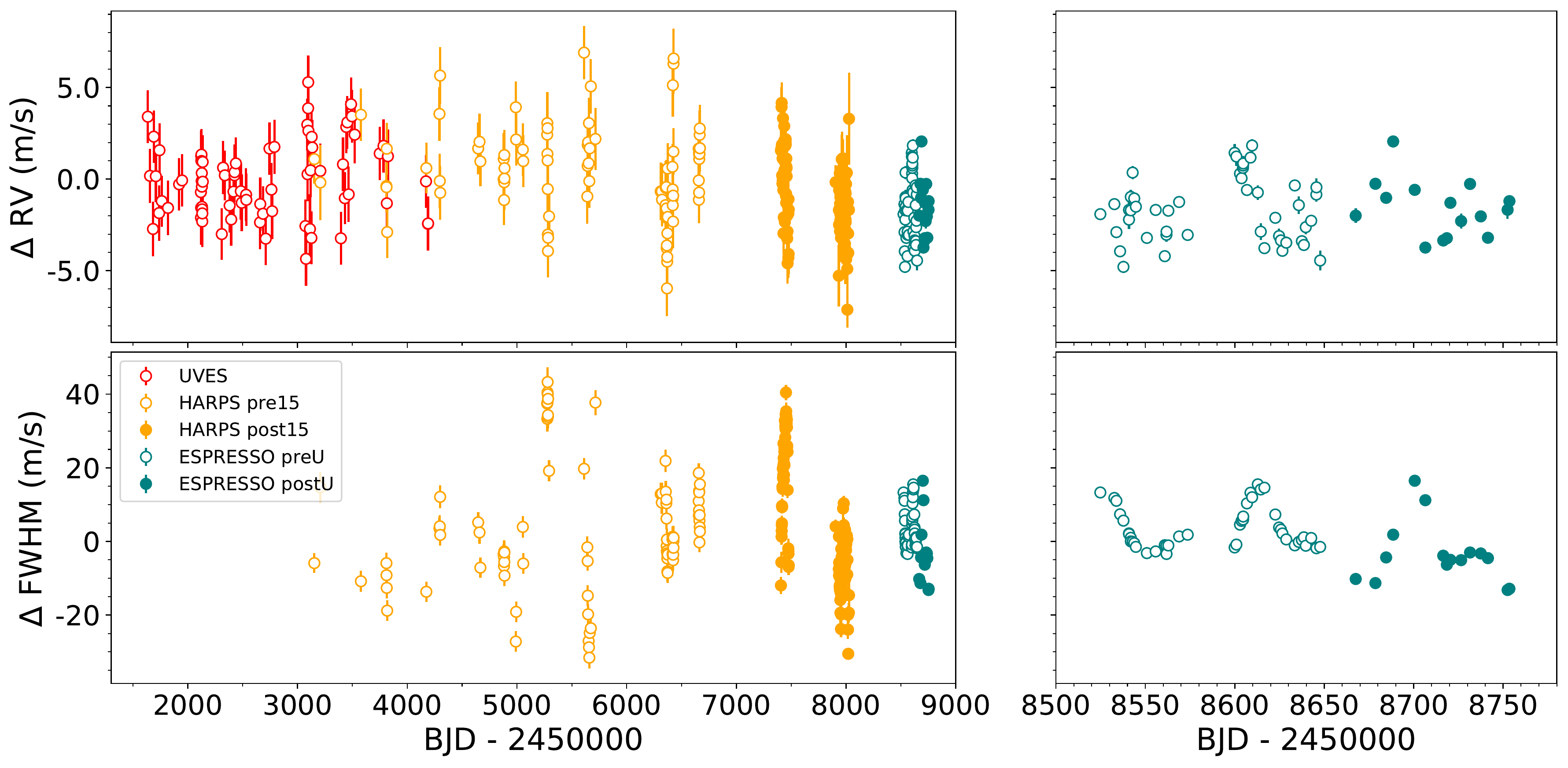}
        \includegraphics[width=18.0cm]{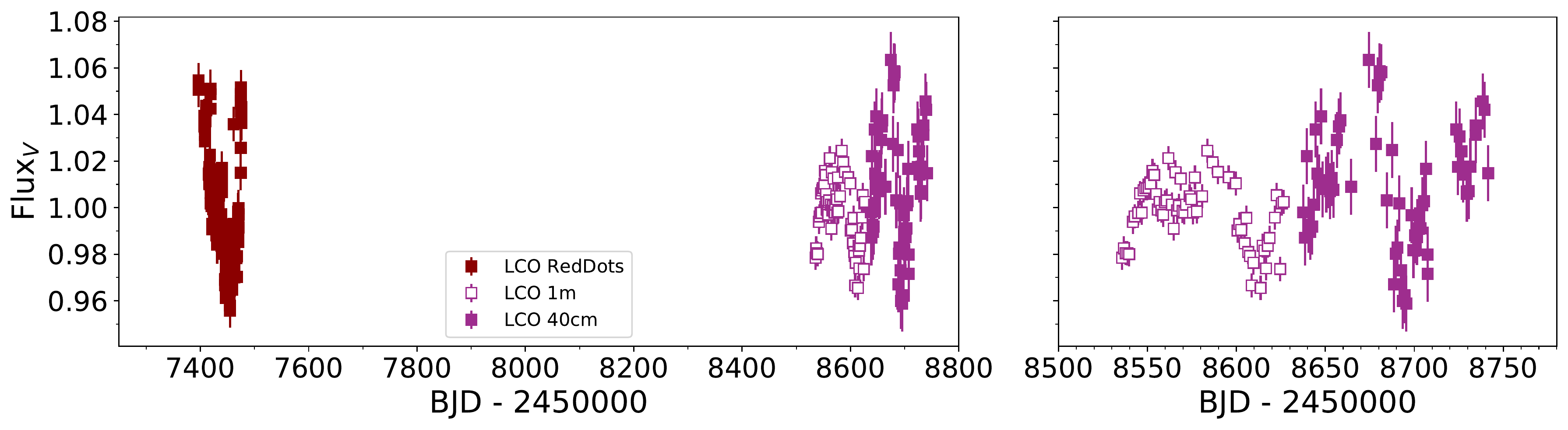}
        \caption{Spectroscopic and photometric data used in this study. The left panels show the complete datasets, while the right panels show a zoom around the 2019 campaign. The y-axis in the right panels uses the same scale as the left panels. }
        \label{data}
\end{figure*}

\subsection{ESPRESSO radial velocity data}

The Échelle SPectrograph for Rocky Exoplanets and Stable Spectroscopic Observations (ESPRESSO) is a fibre-fed high resolution echelle spectrograph installed at the Very Large Telescope (VLT) telescope array in the European Southern Observatory (ESO) Paranal Observatory (Chile) \citep{Pepe2013, GonzalezHernandez2018}. The instrument has a resolving power of approximately $R\sim 140\,000$ over a spectral range from $\sim$380 to $\sim$788 nm and has been designed to attain a long-term RV precision of 10 cm$\cdot$s$^{-1}$. It is contained in a vacuum vessel to avoid spectral drifts due to temperature and air pressure variations, thus  ensuring its stability. Observations were carried out using the Fabry Perot (FP) as simultaneous calibration. The FP offers the possibility of monitoring the instrumental drift with a precision better than 10 cm$\cdot$s$^{-1}$ without the risk of contamination of the stellar spectra by ThAr saturated lines \citep{Wildi2010}. ESPRESSO can be used on any VLT unit telescope (UT), allowing for an efficient observation and permitting a high-cadence observation even on Paranal. More information can be found in the ESPRESSO user manual.\footnote{https://www.eso.org/sci/facilities/paranal/instruments/espresso/doc.html.}

We obtained 67 individual spectra  as part of the ESPRESSO GTO, as part of programme ID 1102.C-744 (PI: F.Pepe). Measurements were taken in ESPRESSO’s 1UT high resolution (HR) mode with 15 minutes of integration time. More information on the different observing modes can be found on the ESO instrument page\footnote{ https://www.eso.org/sci/facilities/paranal/instruments/espresso.html.}. ESPRESSO is equipped with its own pipeline (Lovis et al. \textit{future work}) providing extracted and wavelength-calibrated spectra, as well as RV measurements. The RV measurements are determined by a Gaussian fit of the cross-correlation function (CCF) of the spectrum with a binary mask computed from a stellar template \citep{Baranne1996, PepeMayor2000}. The mask was created using an ESPRESSO spectrum of Proxima as a template. Lines were identified through an automatic line-searching algorithm based on the spectrum derivative.  The pipeline, version 2.0.0, is fully available to download from the ESO pipeline website \footnote{http://eso.org/sci/software/pipelines/.}. 

We removed one data point, with a photon noise $\sim$30 m$\cdot$ s$^{-1}$, taken during a night with poor weather conditions. Another three spectra are discarded due to a flare event during the observations. While it has been shown that flare events do not affect the RV measurements to a level of $\sim$10 m$\cdot$ s$^{-1}$ \citep{Reiners2009}, that is not necessarily the case for sub m$\cdot$ s$^{-1}$ precision. The RV measurements taken during the flare event show a significant correlation with the level of emission in the core of chromospheric lines, and therefore we opted not to use them. In order to validate the results, we tested that the inclusion of those points did not significantly alter the results. The flare was identified by measuring higher than usual flux in several chromospheric emission lines. This leads to 63 RV measurements with a dispersion of 1.9 m$\cdot$s$^{-1}$ and a median RV noise of 0.26 m$\cdot$s$^{-1}$, spread along a baseline of seven months. Along with the velocity, we obtain a measurement of the full width half maximum (FWHM) of the CCF, which will be used for the characterisation of stellar activity.  We opted not to use the bisector, since the amount of information it provides depends on the v$\cdot$sin \textit{i} and has proved to be uninformative for slow rotating small stars, like Proxima \citep{Saar1997, Bonfils2007}. Taking advantage of the extended wavelength range of ESPRESSO and the collecting power of the VLT, we also created a set of chromatic velocities that we used to study the wavelength dependence of the RVs of Proxima. ESPRESSO underwent an intervention during June 2019 to update its fiber-link. This resulted in improved efficiency, with a gain of $\sim$50\% flux, but with the current data it is still unclear whether or not it introduced an RV offset between the data obtained before and after the intervention (Pepe et al. \textit{A\&A submitted}).  

\subsection{Radial velocity data in the literature}

In combination with the ESPRESSO data we include the measurements taken between 2003 and 2017 with the High Accuracy Radial velocity Planet Searcher (HARPS) spectrograph \citep{Mayor2003}. These data where obtained by the Geneva/Grenoble survey \citep{Bonfils2013} and the RedDots project \citep{AngladaEscude2016}\footnote{https://reddots.space/}, under programmes 072.C-0488, 082.C-0718, 183.C-0437, 191.C-0505, 096.C-0082, 099.C-0205, and 099.C-0880. This dataset is comprised of 196 individual measurements that include two high cadence campaigns in 2016 and 2017. The RVs were calculated using the TERRA package \citep{AngladaEscude2012} and have been obtained from \citet{Damasso2020}. In 2015 HARPS was updated with new fibres, which improved its stability but also caused an RV offset with respect to previous measurements \citep{LoCurto2015}. For this reason we treat both HARPS datasets independently. The majority of the data were obtained without simultaneous calibration, which limits the stability of HARPS to a level of $\sim$1 m$\cdot$ s$^{-1}$. 

Along with the previous data we include the RVs taken with the Ultraviolet and Visual Echelle Spectrograph (UVES,  \citet{Dekker2000}) and analysed in \citet{Damasso2020}. This dataset consists of 77 nightly binned UVES RVs obtained between 2000 and 2007. The UVES data were obtained in one of the early RV surveys for planets around M-dwarfs under ESO programme IDs: 65.L-0428, 66.C-0446, 267.C-5700, 68.C-0415, 69.C-0722, 70.C-0044, 71.C-0498, 072.C-0495, 173.C-0606, and 078.C-0829 (PI: M. Kürster). The data reduction and RV measurement is decribed in \citet{Butler2019}. The UVES data do not include a measurement of the FWHM, as it is not easily available due to calibration by the iodine gas absorption cell.

In Fig.~\ref{data} the top and mid panels show the RV and FWHM data used in this study. ESPRESSO data prior to the fiber link upgrade have been labelled \textit{preU}, while the data after the upgrade are labelled \textit{postU}. For the case of HARPS, the data before the 2015 fibre upgrade have been labelled \textit{pre15} and the data after the 2015 fibre upgrade \textit{post15}. Figure~\ref{data} and all figures thereafter were prepared using \texttt{Matplotlib} \citep{Matplotlib}. 

\subsection{Photometric data}

In order to have a continuous monitoring of the stellar activity of the star, Proxima was observed with the Las Cumbres Observatory (LCO, \citet{Brown2013}) 0.4 m, using the SBIC STL-6303 cameras, and 1.0 m telescopes, using the SINISTRO cameras on Cerro Tololo, from February 20 2019 to September 14 2019. The target was observed daily (if conditions were favourable) in $V$ band. From February 20 2019 to May 27 2019 we used the 1.0 m telescopes, taking a time series of 11 images with an exposure time of 20 seconds per frame approximately ten  minutes considering overheads). From June 01 2019 to September 14 2019 we used the 0.4 m telescopes to obtain 18 images with an exposure time of 45 seconds ($\sim$15 minutes considering overheads) per visit. The raw images were reduced by LCO's pipeline \texttt{BANZAI} and aperture photometry was performed on the calibrated images using \texttt{AstroImageJ} \citep{Collins2017}. For each night a fixed circular aperture was selected by \texttt{AstroImageJ} and aperture photometry was performed using this aperture on the target and a set of five reference stars. This set is the same for the 40-cm and 1-m telescopes time series. We obtained 65 nightly V-band measurements using the 1-m telescope, and 62 measurements using the 40-cm telescopes. We combined this data with the available LCO data from the RedDots campaign (170 measurements, also obtained from the RedDots website).  We obtained a typical precision of 0.5\% in relative flux with the 1 m telescope measurement, and 1.2\% in relative flux with the 0.4m telescope measurements. Figure~\ref{data} shows the photometric data used in this study.

\section{Confirmation of Proxima b} \label{sec:prox_b}

Proxima b was detected by combining 147 HARPS nightly binned RV measurements and 77 nightly binned UVES measurements \citep{AngladaEscude2016}, with photon noise uncertainties ranging from 1 to 2 m$\cdot$s$^{-1}$. The planet was characterised as orbiting with a period of 11.186 $\pm$ 0.002 days and as having a mass of 1.27 $\pm$ 0.19 M$_{\oplus}$. It showed a semi-amplitude of 1.38 $\pm$ 0.21 m$\cdot$s$^{-1}$, comparable to the typical error bars of the individual measurements, and a poorly constrained eccentricity.

Proxima shows stellar-induced RV variations larger than the amplitude assigned to Proxima b. Modelling these activity-induced RV variations is key for the proper extraction of the planetary parameters. The Gaussian processes (GP; see \citet{Rasmussen2006} and \citet{Roberts2012}) framework has become one of the most successful methods in the analysis of stellar activity in RV time series (e.g. \citet{Haywood2014}).The stellar noise is described by a covariance functional form and the parameters attempt to describe the physical phenomena to be modelled. The GP framework can be used to characterise the activity signal without requiring a detailed knowledge of the distribution of active regions on the stellar surface, their lifetime, or their temperature contrast. Recently \citet{Damasso2017} showed that the framework could successfully describe the activity pattern of Proxima. We relied on a Gaussian processes model using the  $celerite$\footnote{https://celerite.readthedocs.io.} package \citep{Foreman-Mackey2017} and adopted the rotation kernel described in the same article for the analysis of stellar rotation:

\begin{equation}
\begin{split}
 k(\tau) = {B \over{2+C}} \cdot e^{-\tau/L}\Bigg[cos({{2\cdot \pi \tau}\over{P_{\rm rot}}}) + (1+C)\Bigg]     +\\
 + (\sigma^2_{RV}(t) + \sigma^2_{j}) \cdot \delta_{\tau}
\end{split}
,\end{equation}

 \noindent where B represents the covariance amplitude, \textit{P}$_{\rm rot}$ is the rotation period, \textit{L} represents the lifetime of the configuration of active regions causing the variations, and \textit{C} is the balance between the periodic and the non-periodic parts. The equation also includes a term of uncorrelated noise ($\sigma$), independent for every instrument, added quadratically to the diagonal of the covariance matrix to account for all unmodelled noise components, such as uncorrected activity or instrumental instabilities. The Kronecker delta function is $\delta_{\tau}$  , and $\tau$ represents an interval between two measurements, $t-t'$. This kernel emulates the behaviour of the classical quasi-periodic kernel by \citet{Haywood2014} with a much smaller computational cost, and has been successfully used to model stellar activity in different scenarios during recent years \citep{Angus2018, Espinoza2019,JonesMatias2019}.  We included independent zero-point velocities and jitter terms for every instrument, including separate terms for HARPS before and after the change of fibre, and for ESPRESSO before and after the intervention performed in June 2019. Along with the GP kernel and the jitter and zero point measurements, we included a linear term for every time series and a Keplerian in the RV dataset to account for the planetary-induced variations. 

We performed a joint model using the FWHM and the RV data together. We used the FWHM as our main activity indicator, as we found it traces well the photometric behaviour of the star. Our model includes a rotation term (GP) and a linear slope, and later we include a Keplerian along with the activity model. The parameters P$_{rot}$, L, and C are shared between the GP components of the RV and FWHM time series, while each one retains a different amplitude.  Section~\ref{sect:activity} provides more detail on the stellar activity of Proxima and the reasons behind our choice of activity proxy. 

\subsection{Modelling the ESPRESSO data} \label{sec:prox_b_esp}

\begin{figure}
        \includegraphics[width=9cm]{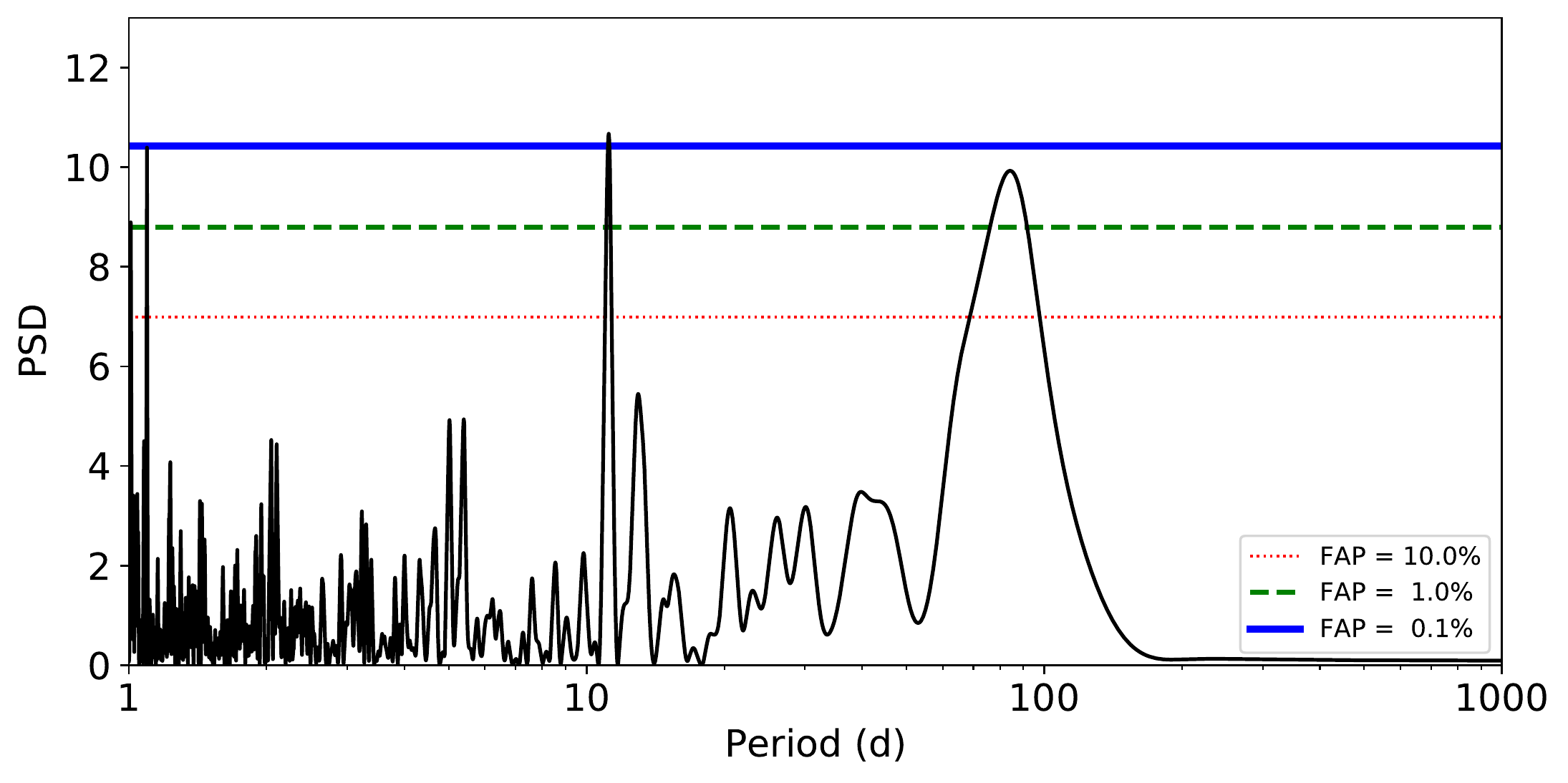}
        \caption{Generalised Lomb-Scargle periodogram of the ESPRESSO RV data.}
        \label{gls_esp}
\end{figure}

We start by modelling the different signals present in the data one by one, akin to a classical pre-whitening on the RVs of ESPRESSO. Figure~\ref{gls_esp} shows the generalised Lomb-Scargle (GLS) periodogram \citep{Zechmeister2009} of the RV data shown in Fig.~\ref{data} after subtracting a linear trend to account for variations longer than our baseline of observations. We use the \textit{PyAstronomy} implementation of the GLS \citep{pyastronomy} with the  periodogram power spectral density (PSD) normalisation and false alarm probability (FAP) levels of \citet{HorneBaliunas1986}. We see two signals that cross the 1\% FAP line. The most prominent of the two corresponds to a signal at 81 days, a period very close to the known rotation period of Proxima, while the second occurs at 11.2 days. There are two extra peaks near the one-day mark that correspond to aliases of the other two signals. 

\begin{figure*}
        \includegraphics[width=18cm]{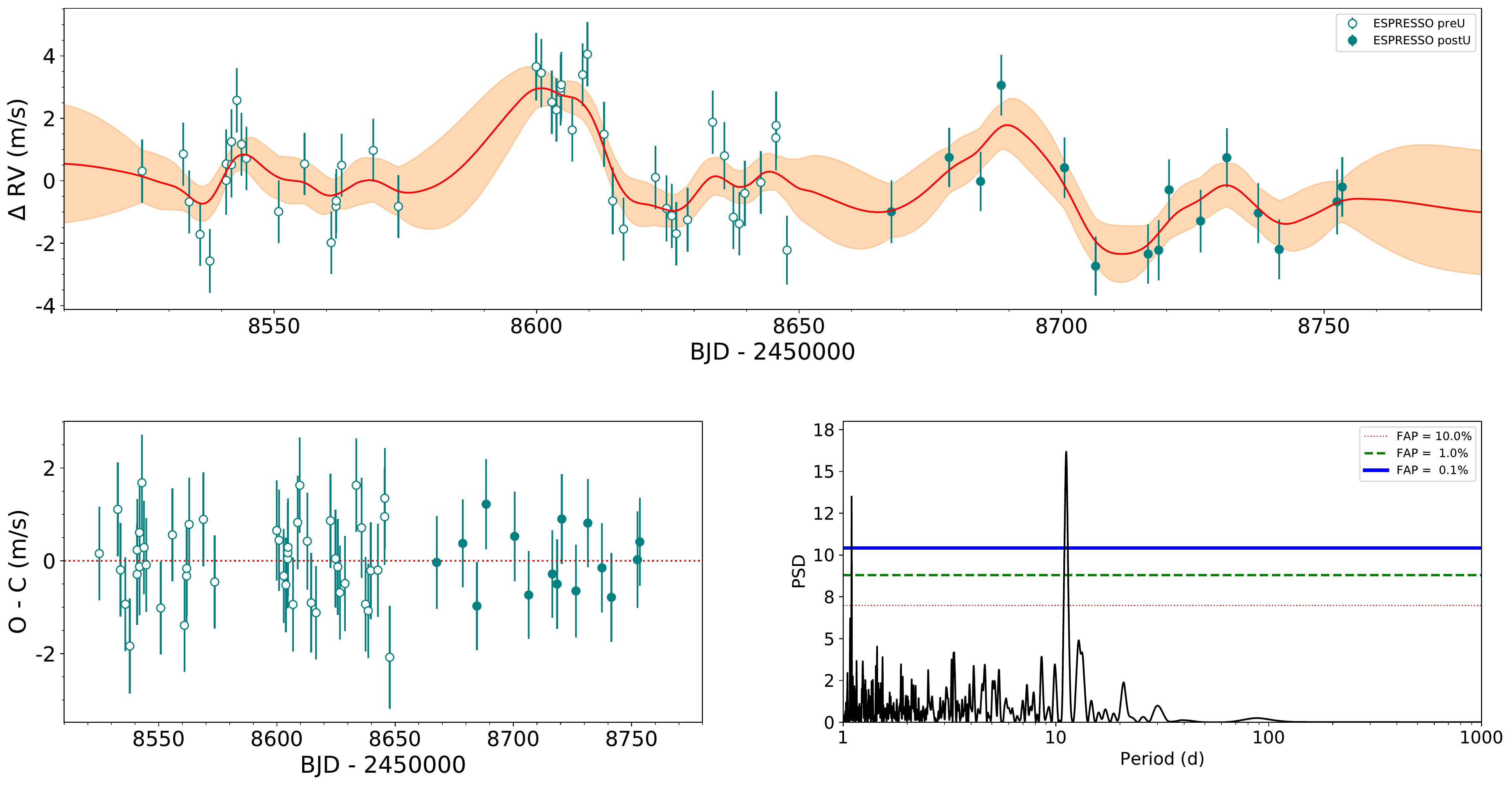}
        \caption{\textbf{Top panel:} ESPRESSO RVs with the best activity-only model. \textbf{Bottom panels:} Detrended ESPRESSO RVs and their periodogram. The error bars in both panels include the jitter component obtained using the activity-only model.}
        \label{rv_timeseries}
\end{figure*}

We apply the described GP model, without any Keplerian component, to the ESPRESSO data by performing Markov chain Monte Carlo (MCMC) simulations using \texttt{emcee} \citep{Foreman-Mackey2013} with a differential evolution algorithm \citep{TerBraak2006,Nelson2014}. This combines the differential evolution (DE) genetic algorithm \citep{Storn1997} into an ensemble MCMC for a more efficient sampling of the parameter space. The model uses separate GP kernels for the FWHM and the RV data, with independent jitter parameters for the data before and after the intervention in ESPRESSO. The parameters \textit{P}$_{\rm rot}$, \textit{L,} and \textit{C} are shared between the four GP kernels. We initialise a number of walkers equal to four times the number of parameters, using uniform ($\mathcal{U}$)  and log uniform ($\mathcal{LU}$)  priors as listed in Table~\ref{tab:1p_parameters_esp}. We run a maximum of 500~000 steps, and then measure the auto-correlation timescale every 5~000 steps to check for convergence. Convergence is assumed when the number of steps is larger than 100 times the auto-correlation timescale, and the timescale changes less than 1$\%$ from the previous measurement, as described in the \texttt{emcee} documentation. The burn-in period is later defined as 20 times the auto-correlation timescale. Table~\ref{tab:1p_parameters_esp} shows the priors used and the final parameters obtained in the MCMC simulations, and Fig.~\ref{post_dist_nop_esp} in Appendix~\ref{ap:post_dist} shows the posterior distribution of the parameters. Using the posterior distribution, we estimate the Bayesian evidence (\textit{logZ}) by following \citet{Perrakis14}. Using only the GP components, we obtain a \textit{logZ} of -259. The GP converges to a rotation period of 51 days, closer to half of the rotation than to a full rotation. This might be caused by the shape of the FWHM variations during this season, combined with the short baseline of observations, of less than three full rotations. The residuals after the fit show a variation slightly below 1 m$\cdot$s$^{-1}$ that is not described by the model. The GLS of the residuals of the data show a very clear and very significant peak at 11.2 days. Figure~\ref{rv_timeseries} shows the ESPRESSO RV data and the detrended data using the stellar activity model described before, and its GLS periodogram. 

\begin{figure*}
        \includegraphics[width=18cm]{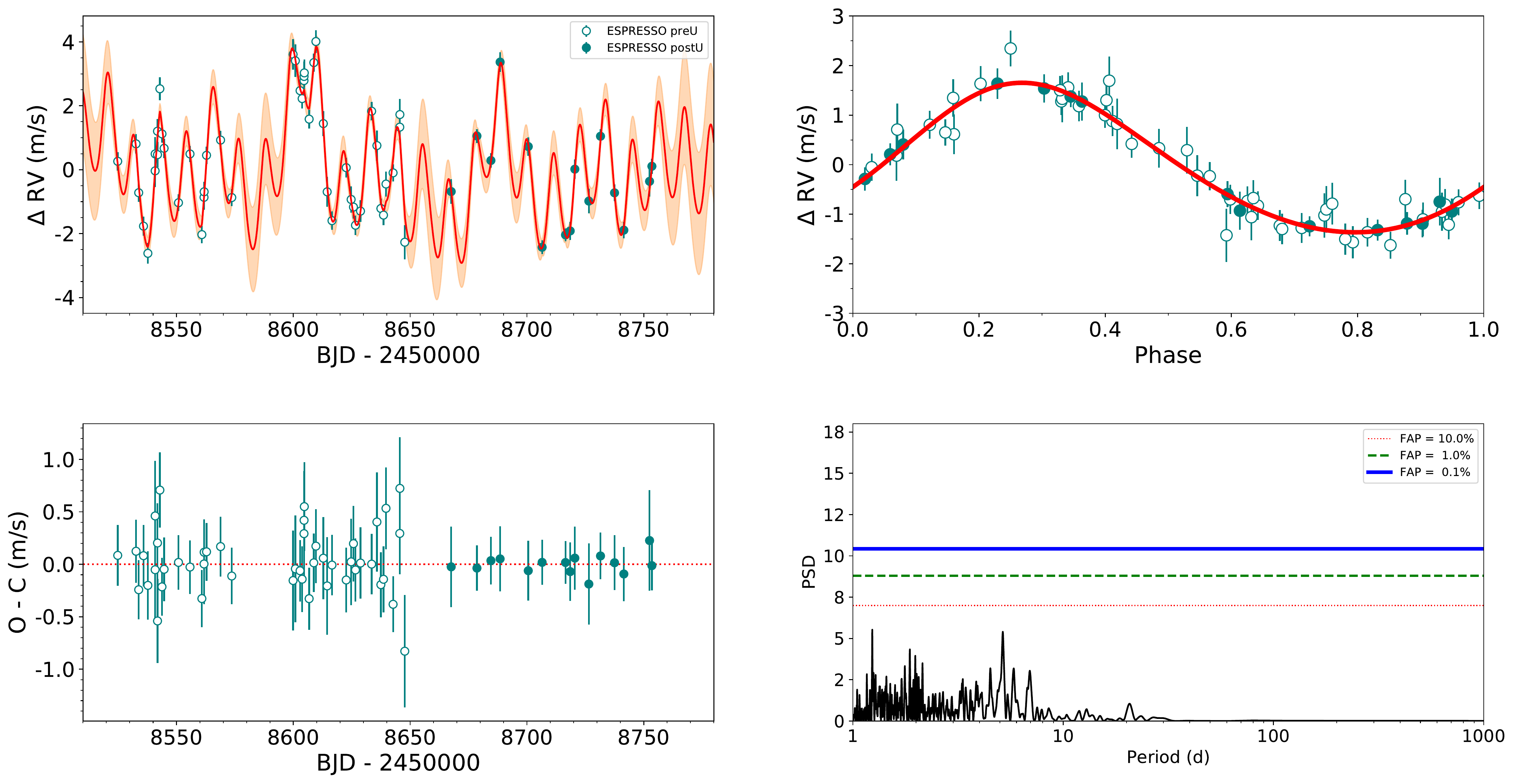}
        \caption{\textbf{Top-left panel:} ESPRESSO RVs along with the best fit using a single planet and the activity model.
        \textbf{Top-right panel:} Phase folded RV curve of the 11.2d signal after subtraction of the GP model component. \textbf{Bottom-left panel:} Residuals after the fit of the full model.  \textbf{Bottom-right panel:} GLS periodogram of the residuals. }
        \label{rv_1p_model}
\end{figure*}

Fitting together the activity model and a Keplerian we obtain a much better fit with significantly smaller residuals, almost exactly at the photon noise level. We run the MCMC with the same configuration as before, considering in this case wide priors for the Keplerian parameters.  Figure~\ref{rv_1p_model} shows the fit together with our RV data. The Keplerian signal converges to a period of 11.22 $\pm$ 0.03 days, with a semi-amplitude of 1.51 $\pm$ 0.15 m$\cdot$s$^{-1}$, and an eccentricity of 0.1. It corresponds to a minimum mass of 1.29 $\pm$ 0.13 M$_{\oplus}$. The rotation component converges once again to half of the true rotation, with a decay timescale slightly longer than the measured periodicity. It shows an amplitude parameter of 5.0 $^{+3.6}_{-1.7}$ (m$\cdot$s$^{-1}$)$^{2}$. As it is a squared amplitude, it translates to a semi-amplitude of $\sim$ 2.2 m$\cdot$s$^{-1}$. This amplitude of the RV-induced signal is compatible with other M-dwarfs with similar levels of chromospheric activity \citep{Masca2018}. The residuals of the data show a dispersion of 0.27 m$\cdot$s$^{-1}$, similar to the average photon-noise-induced RV error, and no significant signals in the periodogram. 

\begin{figure}
        \includegraphics[width=9cm]{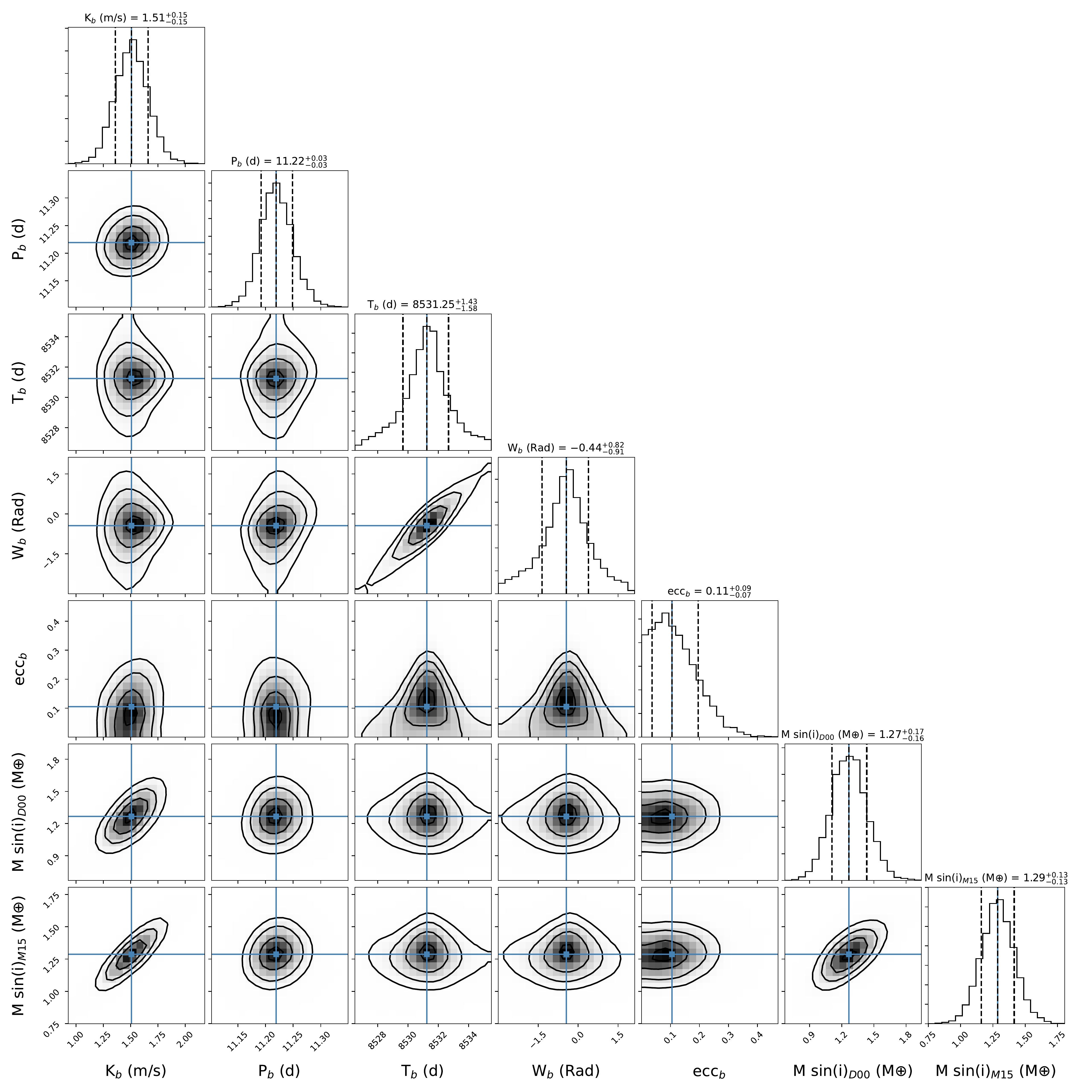}
        \caption{Posterior distributions of the parameters of the Keplerian coming from the joint model of the FWHM and the RV data using the ESPRESSO dataset. Figure~\ref{post_dist_1p_esp} in Appendix~\ref{ap:post_dist} shows the full posterior distribution for all parameters. The figure is made using the \texttt{corner} package \citep{cornerplot}.}
        \label{post_dist_kep_esp}
\end{figure}

Table~\ref{tab:1p_parameters_esp} shows the final parameters obtained and the priors used in the MCMC simulations, and Fig.~\ref{post_dist_kep_esp} shows the posterior distribution of the planetary parameters. Using the GP with a Keplerian signal, we obtain a \textit{logZ} of -219. We measure a $\Delta$ \textit{logZ} of 40, consistent across multiple tries, between the 1-planet and 0-planet solutions. According to the Jeffreys’ scale, this means the 1-planet model is very strongly favoured over the 0-planets model. The full posterior distribution can be found in Fig.~\ref{post_dist_1p_esp}, in Appendix~\ref{ap:post_dist}.

\begin{figure}
        \includegraphics[width=9cm]{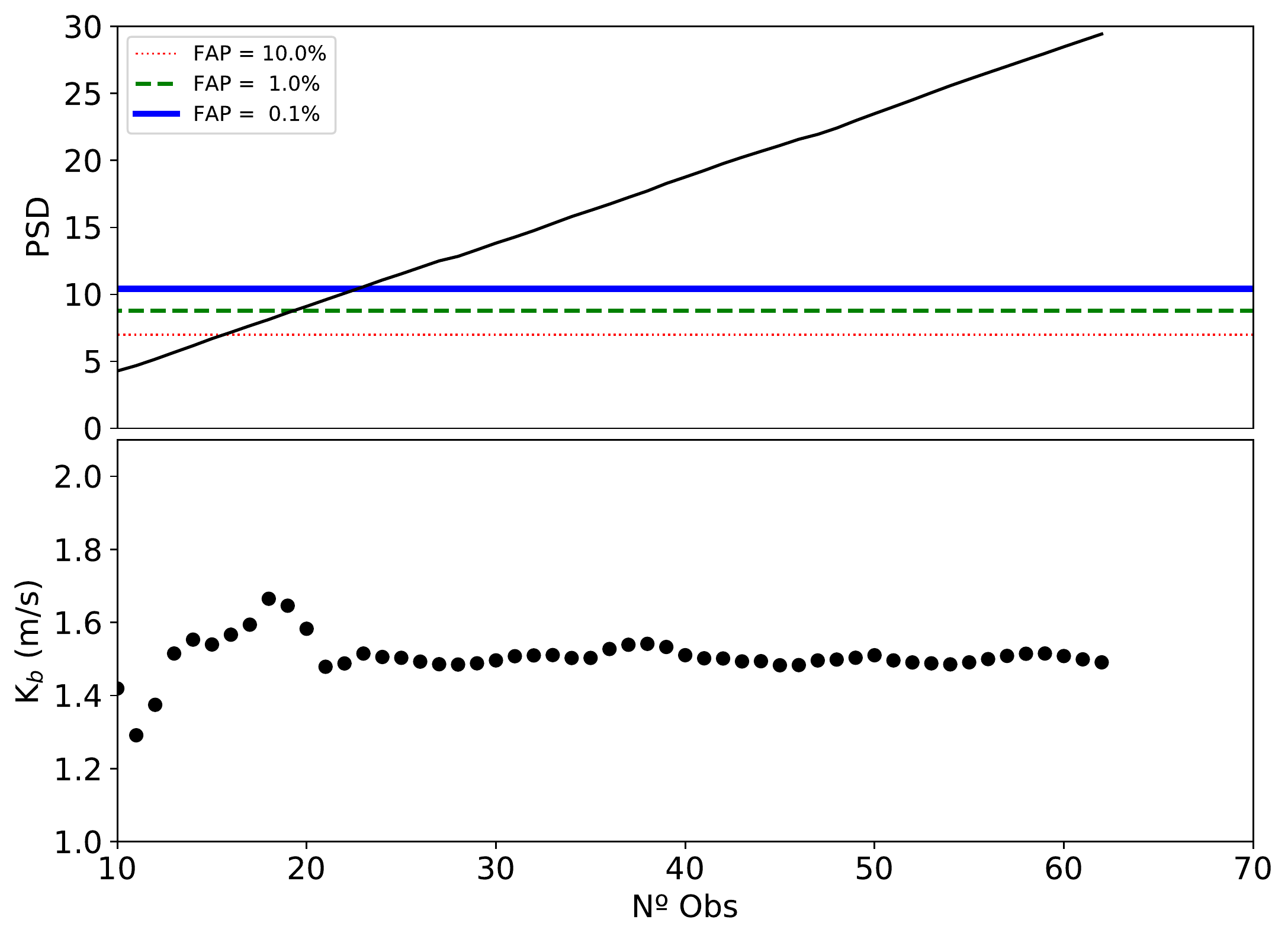}
        \caption{\textbf{Top panel:} Evolution of the periodogram power of the 11.2d signal as a function of the number of observations for the ESPRESSO dataset. \textbf{Bottom panel:} Evolution of the amplitude of the 11.2d signal as a function of the number of observations.}
        \label{stack_1p}
\end{figure}

Following ~\citet{Mortier2017} and ~\citet{Masca2018a} we study the behaviour of the 11.2d signal as a function of the number of measurements and measure the evolution of its amplitude. Keplerian signals are expected to be stable in RV datasets, while activity signals are not, due to their lack of long-term coherency \citep{Masca2018a}. If the 11.2d signal corresponds to a planet, we expect a steady increase in the periodogram power with the number of measurements. Figure~\ref{stack_1p} shows the evolution of the periodogram power of the 11.2d signal shown in Fig.~\ref{rv_1p_model}. The PSD of the periodogram show an almost perfectly linear increase with the number of observations. We also see that after $\sim$20 observations the amplitude of the signal remains stable for the rest of the campaign. It should be noted that the PSD of the signal grows higher than what Fig.~\ref{rv_timeseries} shows. This is caused by the different approach to modelling. While in Fig.~\ref{rv_timeseries} we used the GP to model the activity and then performed the periodogram of the residuals, here we are modelling the Keplerian along with the activity using least-squares minimisation and then subtracting only the activity component to ensure we isolate the 11.2d signal. The bottom panel of the figure shows the measured amplitude coming from the combined model. From very early on the amplitude of the signal stays consistent within error bars. 

\begin{figure*}
        \includegraphics[width=18cm]{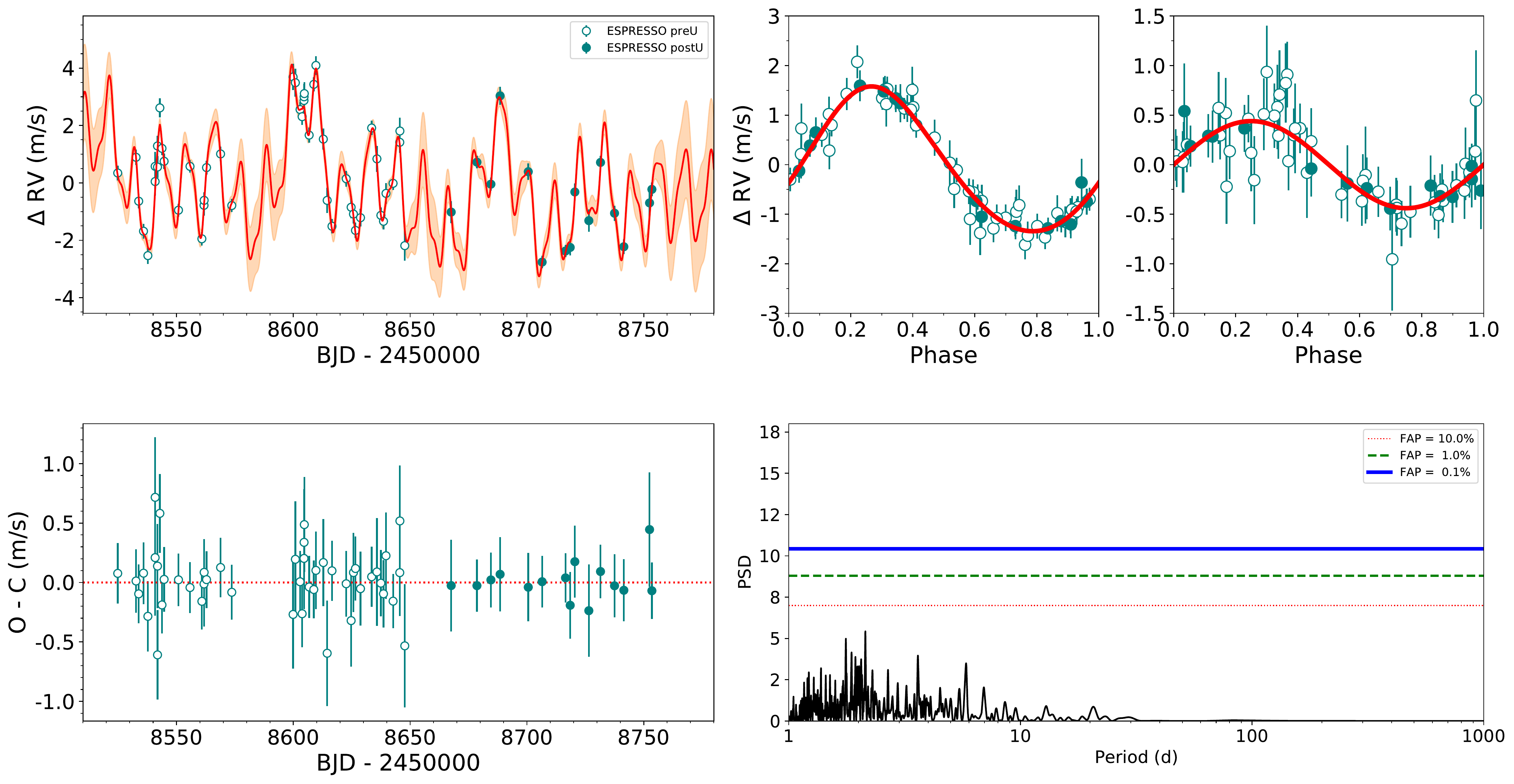}
        \caption{\textbf{Top-left panel:} ESPRESSO RVs along with the best fit using a Keplerian, a sinusoidal, and the activity model.
        \textbf{Top-right panel:} Phase folded RV curve of the 11.2d and 5.15d signals. \textbf{Bottom-left panel:} Residuals after the fit of the full model.  \textbf{Bottom-right panel:} GLS periodogram of the residuals. }
        \label{rv_2p_model}
\end{figure*}

\begin{figure}
        \includegraphics[width=9cm]{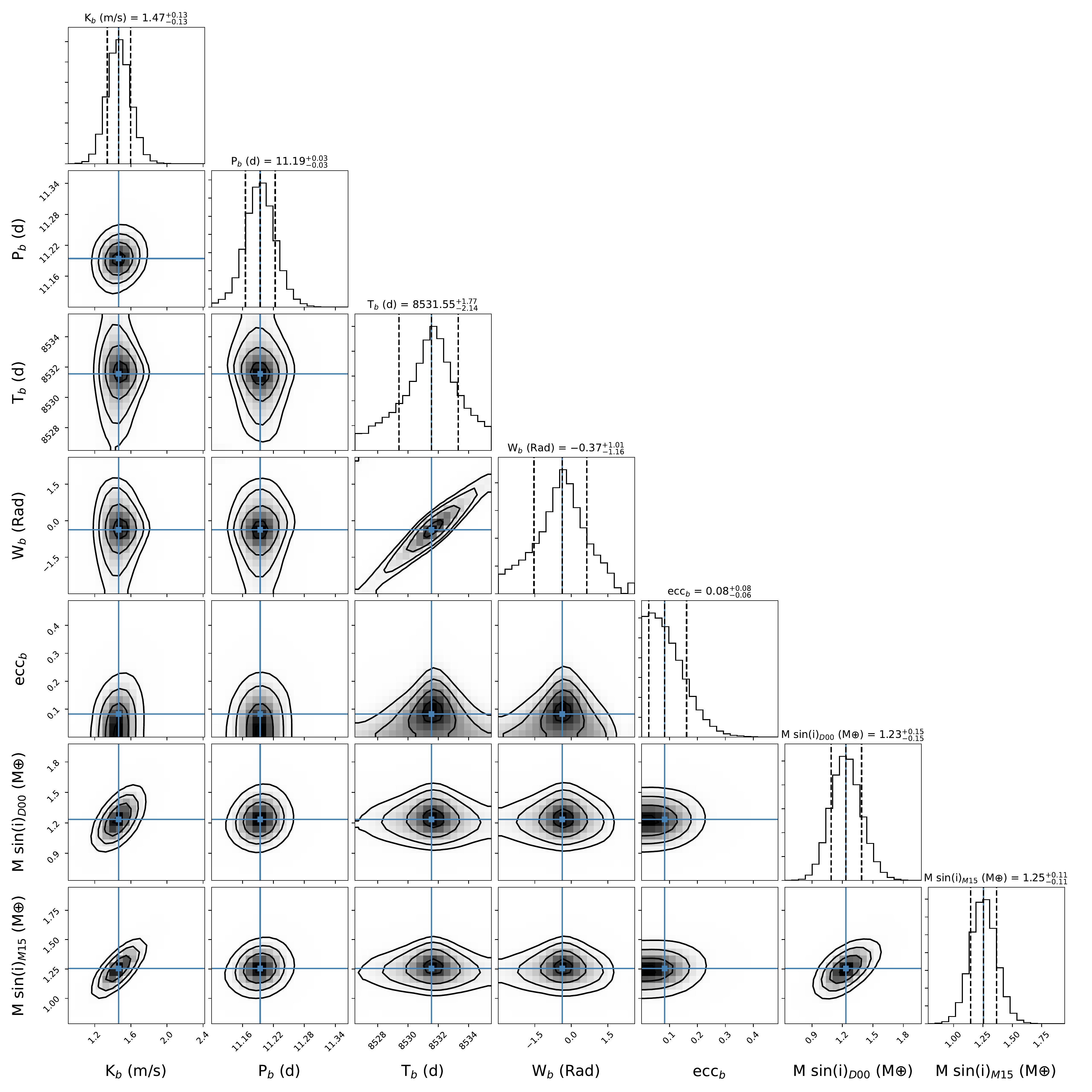}
        \includegraphics[width=9cm]{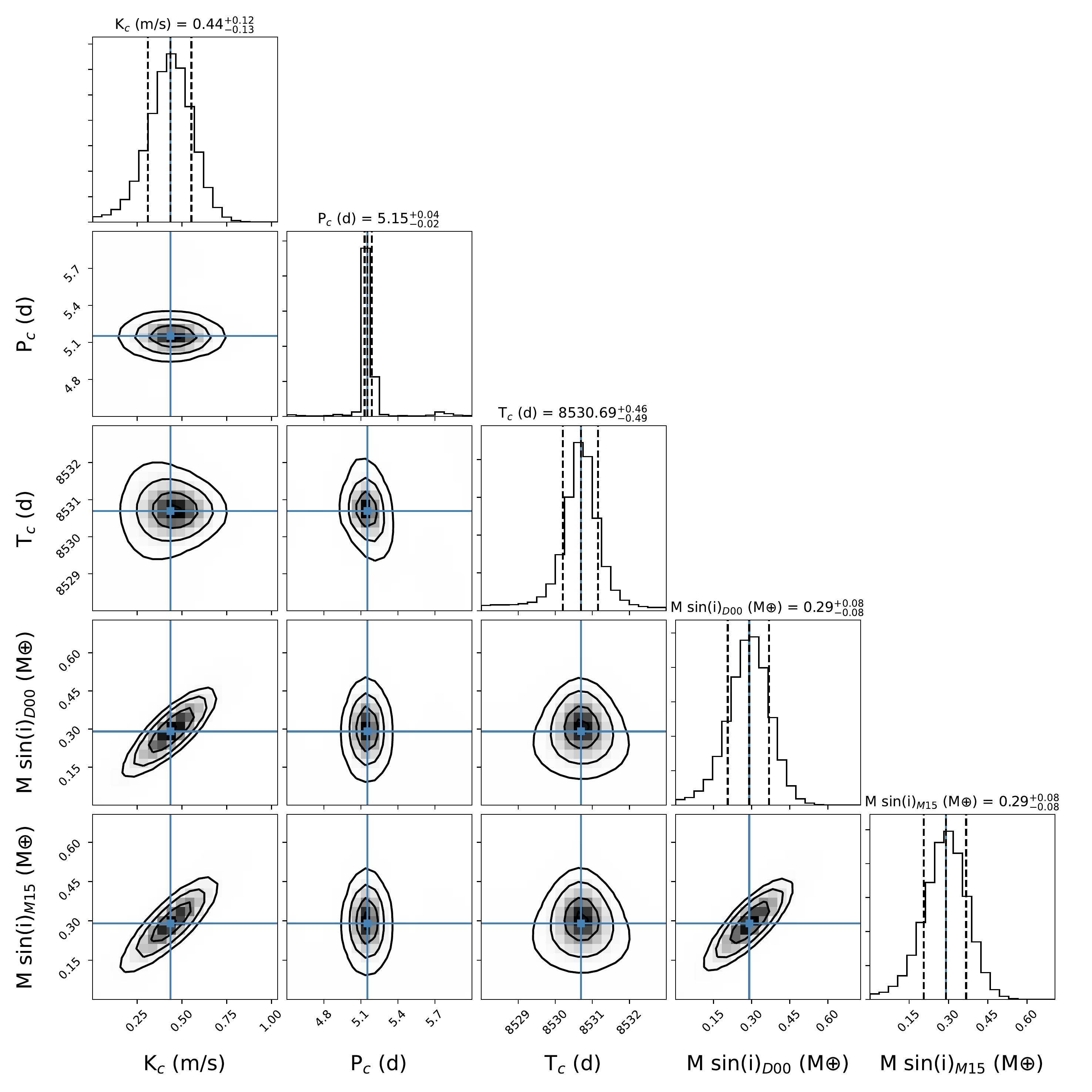}
        \caption{Posterior distributions of the parameters of the Keplerian and sinusoidal components, coming from the joint model of the FWHM and the RV data using the ESPRESSO dataset. Figure~\ref{post_dist_2p_esp} in Appendix~\ref{ap:post_dist} shows the full posterior distribution for all parameters.}
        \label{post_dist_kep_esp_2p}
\end{figure}

Although the periodogram of the residual does not show any significant signal, it is important to acknowledge that a GP and uncorrelated noise components may be absorbing additional low amplitude signals present in the data. With this in mind, we run a last simulation including an extra sinusoidal to account for a hypothetical extra planet. Using the same prior as for the first planet (2-20 days), we obtain a power excess at 5.15 days with a semi-amplitude of 0.44 $\pm$ 0.13 m$\cdot$s$^{-1}$, the same as was seen before in the GLS of the residuals. Figure~\ref{rv_2p_model} shows the best fit of the data using this model, while Table~\ref{tab:1p_parameters_esp} shows the final parameters obtained and the priors used in the MCMC simulations, and Fig.~\ref{post_dist_kep_esp_2p} shows the posterior distribution of the two signals. We measure a $\Delta$ \textit{logZ} of 7, which would make this signal significant according to many interpretations of the Bayesian evidence. This signal, if caused by a planet, would correspond to a minimum mass of 0.29 $\pm$ 0.08 M$_{\oplus}$ at an orbital distance of 0.02895 $\pm$ 0.00022 AU. The equilibrium temperature of the planet candidate would be 330 $\pm$ 30 K for a Bond albedo A = 0.3. However the planetary origin of the signal is far from guaranteed. Though the posterior distributions and the fit look good, the model is only barely significant over the 1-planet model, and no significant peaks at this period show up in the previous periodograms.  Section~\ref{sect:activity} shows an inconsistent presence of the signal across the ESPRESSO wavelengths, which casts doubts on its planetary origin. The lack of long-term ESPRESSO coverage also makes it impossible to test if the signal can be reliably detected in different seasons, with different levels of stellar activity. More data would be needed to confirm the true nature of this 5.15 days signal.

\begin{table*}
\begin{center}
\caption{Parameters of the global model using exclusively ESPRESSO data. Mass $\cdot$ sin(i)$_{D00}$ shows the minimum mass of Proxima b using the \citet{Delfosse2000} estimation for the mass of Proxima, and  Mass $\cdot$ sin(i)$_{M15}$ shows the minimum mass using the \citet{Mann2015} estimation. \label{tab:1p_parameters_esp}}
\begin{tabular}[center]{l c c c  c c}
\hline
Parameter  & Priors & GP & GP + 1 Signal & GP + 2 Signals \\ \hline

\textit{Keplerian} \\
Period [days] &  
$\mathcal{U}$ (2 , 20) &  & 11.219$^{+0.029}_{-0.027}$ &  11.194$^{+0.029}_{-0.028}$ \\
K [m$\cdot$s$^{-1}$] &  $\mathcal{U}$ (0 , 5) & & 1.51$^{+0.15}_{-0.15}$ &  1.47$^{+0.13}_{-0.13}$\\
e & $\mathcal{U}$ (0 , 1) & & 0.105$^{+0.091}_{-0.068}$ & 0.082$^{+0.079}_{-0.056}$ \\
$T_{P} - 2450000$ [d] & $\mathcal{U}$ (8525 , 8535) & & 8531.3 $^{+1.4}_{-1.6}$ & 8531.5 $^{+1.8}_{-2.1}$  \\
$\omega$ [rad] & $\mathcal{U}$ (-$\pi$ , $\pi$)& & --0.44 $^{+0.82}_{-0.91}$  & --0.4 $^{+1.0}_{-1.2}$   \\ \\
\textit{Derived parameters} \\
Semi-major axis [AU] & & & 0.04864 $^{+0.00031}_{-0.00031}$ & 0.04858 $^{+0.00031}_{-0.00031}$  \\
Mass $\cdot$ sin(i)$_{D00}$ [M$_{\oplus}$] &&& 1.27 $^{+0.17}_{-0.16}$ & 1.23 $^{+0.15}_{-0.15}$ \\
Mass $\cdot$ sin(i)$_{M15}$ [M$_{\oplus}$] &&& 1.29 $^{+0.13}_{-0.13}$ & 1.25 $^{+0.11}_{-0.11}$ \\
\\
\textit{Sinusoidal} \\
Period [days] &  
$\mathcal{U}$ (3 , 10) &  && 5.152$^{+0.036}_{-0.023}$ \\
K [m$\cdot$s$^{-1}$] &  $\mathcal{U}$ (0 , 5) & && 0.44$^{+0.12}_{-0.13}$ \\
$T_{0} - 2450000$ [d] & $\mathcal{U}$ (8525 , 8535) & && 8530.69$^{+0.46}_{-0.49}$ \\
 \\
\textit{Activity model} \\
B$_{GP}$ RV [(ms$^{-1})^{2}$] & $\mathcal{LU}$ (0.1 , 50) &  5.1 $^{+554}_{-2.8}$ &  5.0 $^{+3.6}_{-1.7}$&  4.0 $^{+2.7}_{-1.5}$\\
B$_{GP}$ FWHM [(ms$^{-1})^{2}$]& $\mathcal{LU}$ (0.1 , 300)&  53 $^{+32}_{-16}$&  51 $^{+28}_{-15}$ &  54 $^{+29}_{-16}$\\
P$_{GP}$ [d] & $\mathcal{LU}$ (30 , 300) & 51.8 $^{+10.4}_{-5.7}$& 47.7 $^{+6.0}_{-4.1}$  & 46.9 $^{+4.8}_{-3.5}$ \\
L$_{GP}$ [d]& $\mathcal{LU}$ (30 , 300) & 64 $^{+44}_{-24}$ & 66 $^{+40}_{-22}$   & 73 $^{+44}_{-24}$  \\
log C$_{GP}$ & $\mathcal{U}$ (-10 , 10) &  -300$^{+210}_{-200}$   &  -290$^{+210}_{-200}$  &  -300$^{+205}_{-210}$  \\ \\
\textit{Noise} \\
Jit$_{E~pre}$ RV [m$\cdot$s$^{-1}$]& $\mathcal{LU}$ (0 , 5) & 0.98$^{+0.21}_{-0.20}$ & 0.15$^{+0.23}_{-0.13}$ & 0.066$^{+0.150}_{-0.047}$ \\ 
Jit$_{E~post}$ RV [m$\cdot$s$^{-1}$]& $\mathcal{LU}$ (0 , 5) & 0.92 $^{+0.41}_{-0.34}$ & 0.073 $^{+0.205}_{-0.041}$ & 0.069$^{+0.208}_{-0.051}$ \\ \\
Jit$_{E~pre}$ FWHM [m$\cdot$s$^{-1}$]& $\mathcal{LU}$ (0 , 5) & 0.063 $^{+0.171}_{-0.045}$ & 0.073 $^{+0.205}_{-0.053}$ & 0.070$^{+0.200}_{-0.052}$\\ 
Jit$_{E~post}$ FWHM [m$\cdot$s$^{-1}$]& $\mathcal{LU}$ (0 , 5) & 0.103 $^{+0.444}_{-0.082}$ & 0.088 $^{+0.374}_{-0.068}$& 0.096$^{+0.377}_{-0.075}$ \\ \\
\textit{Polynomial terms} \\
V0$_{E~pre}$ RV [m$\cdot$s$^{-1}$]& $\mathcal{U}$ (-10 , 10) & -0.9 $^{+1.3}_{-1.4}$ & -0.9 $^{+1.3}_{-1.4}$ & -0.9 $^{+1.2}_{-1.2}$\\ 
V0$_{E~post}$ RV [m$\cdot$s$^{-1}$]& $\mathcal{U}$ (-10 , 10) & 2.2 $^{+2.1}_{-2.0}$ & 1.9 $^{+1.9}_{-2.0}$& 2.1 $^{+1.7}_{-1.7}$\\ 
Lin  RV [m$\cdot$s$^{-1}$ $^{-d}$]& $\mathcal{U}$ (-0.1 , 0.1) & -0.005 $^{+0.017}_{-0.018}$ & -0.002 $^{+0.016}_{-0.016}$ & -0.005 $^{+0.014}_{-0.014}$ \\ \\
V0$_{E~pre}$ FWHM [m$\cdot$s$^{-1}$]& $\mathcal{U}$ (-10 , 10) & 1.6 $^{+4.5}_{-4.2}$ & 1.5 $^{+4.2}_{-4.3}$& 1.5 $^{+4.4}_{-4.4}$\\ 
V0$_{E~post}$ FWHM [m$\cdot$s$^{-1}$]& $\mathcal{U}$ (-10 , 10) & 3.8 $^{+6.3}_{-6.5}$  & 3.6 $^{+6.1}_{-6.2}$& 3.5 $^{+6.3}_{-6.4}$  \\ 
Lin  FWHM [m$\cdot$s$^{-1}$ $^{-d}$]& $\mathcal{U}$ (-0.1 , 0.1) & -0.053 $^{+0.052}_{-0.051}$& -0.054 $^{+0.050}_{-0.050}$ & -0.054 $^{+0.049}_{-0.049}$ \\ \\
\textit{Residuals} \\
RMS res [m$\cdot$s$^{-1}$] && 0.82 & 0.27 & 0.26 \\ \\
$log~Z$ & & -259 & -219 & -212 \\\hline
\end{tabular}
\end{center}
\end{table*}

Proxima b, as measured by the ESPRESSO data, shows a minimum mass of 1.29 $^{+0.13}_{-0.13}$ M$_{\oplus}$ when using the \citep{Mann2015} mass estimation for Proxima, or 1.27 $^{+0.17}_{-0.16}$ when using the \citet{Delfosse2000} estimation. The result is compatible with the measurements of \citet{AngladaEscude2016} and \citet{Damasso2017}. Originally the eccentricity was constrained only by an upper limit of 0.35, which \citet{Damasso2017} revised to  0.17$^{+0.21}_{-0.12}$. Our measurement points to a smaller eccentricity, of  0.105$^{+0.091}_{-0.068}$, compatible with zero at less than 2$\sigma$. We obtain slightly different period and amplitude measurements from the measurement obtained by \citet{AngladaEscude2016}, but given the error bars they remain 1-sigma compatible.

It is worth noting that at $\sim$ 27 cm$\cdot$s$^{-1}$ photon-noise-induced RV error, the residuals of the ESPRESSO RV measurements show a dispersion at a very similar level. This suggests that at least to this precision the instrumental stability is not the limiting factor of the measurements. That is reflected by the jitter measurements we obtain for ESPRESSO, which are very close to zero. Given this result, it seems that the photon noise of the data is our limiting factor. We selected 15 minutes of integration time as a compromise between time invested and signal to noise. With longer integration times, or co-adding consecutive spectra, it is possible to improve the precision of the measurements even around an M-dwarf like Proxima to survey for smaller mass planets. We cannot rule out the possibility that the GP kernel is absorbing RV variations that could otherwise be attributed to instrumental instabilities. Gaussian processes can, in some cases, over-fit the data \citep{Feng2016}, although it does not seem to be the case with the other instruments, as seen in Sect.~\ref{sec:prox_b_full}, so we do not expect the GP to treat the ESPRESSO data in a different way. The ESPRESSO data on its own provides an independent and significant detection that supports the presence of a planet orbiting with a period of 11.2 days around Proxima, further confirming the claim by \citet{AngladaEscude2016}.

\subsection{Including HARPS and UVES archival data} \label{sec:prox_b_full}

The HARPS+UVES archival data are comprised of 274 observations spread across 15 years. While the expected instrumental noise of the measurements is much higher than that of the ESPRESSO measurements, they can provide an invaluable assistance in tightening the parameters of the model. 

We re-ran the MCMC analysis using the combined UVES+HARPS+ESPRESSO dataset. The model is the same as described before, with the same configuration for the MCMC. We tested again a model based only on the GP component, and models with one and two extra signals. The priors used are the same as used for the ESPRESSO-only analysis (see Table~\ref{tab:1p_parameters}). 

Re-testing the presence of the recently published Proxima c \citep{Damasso2020} is beyond the scope of this paper, as the ESPRESSO data cannot offer any insights yet due to its short baseline of observations when compared to the long period of the planet candidate (1907 d). A much longer running programme would be needed for ESPRESSO to add meaningful constraints to the presence of Proxima c. We expect any effect that the presence of a long period and low amplitude signal could have in our data to be absorbed by the GP component. 

As with the previous simulations, we set a number of walkers equal to four times the number of parameters, and run the chains for up to 500~000 steps. We found the model with one planet to be much more significant than the pure-activity model ($\Delta$\textit{logZ} $\sim$66). Table~\ref{tab:1p_parameters} shows the final parameters obtained and the priors used in the MCMC simulations. Although HARPS and UVES lack the required precision, we tried to detect the presence of the 5.15 days signal. As with the ESPRESSO data alone, we found some evidence for the presence of the signal, with an amplitude of 35 cm$\cdot$s$^{-1}$ in this case. Figure~\ref{rv_full_dataset_2p} shows the best fit to the data using this mode. We found the $\Delta$\textit{logZ} to be still~seven, the same as using the ESPRESSO data on its own. The use of the archival data did not provide extra evidence for the presence of this signal. This is not unexpected, as it shows that the archival data are not precise enough to help in confirming the nature of this signal. Future efforts with ESPRESSO would be needed to obtain a definitive answer. 

\begin{figure*}
        \includegraphics[width=18cm]{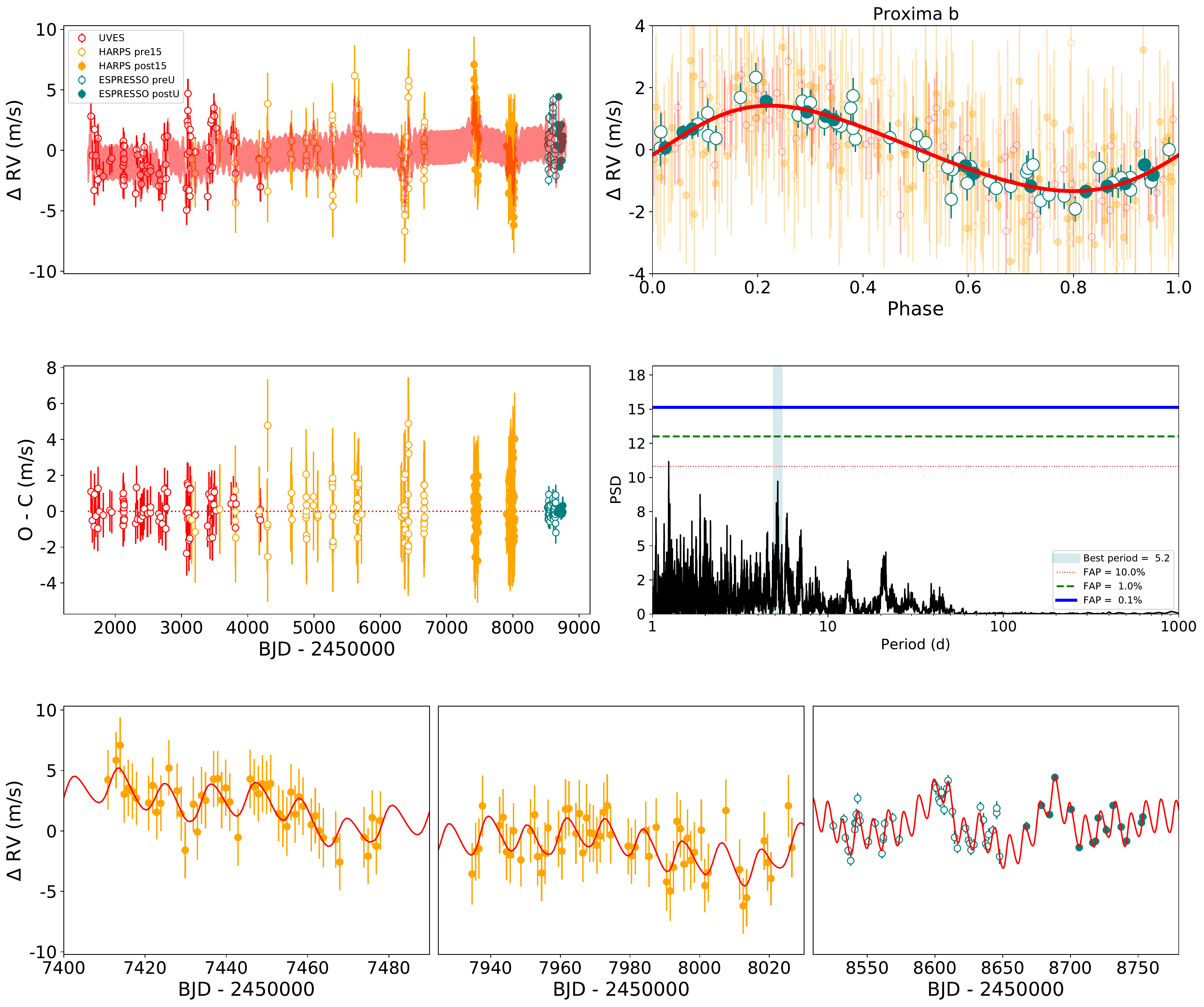}
        \caption{\textbf{Top panels:} Full RV dataset along with the best fit using a Keplerian and the activity model, and phase folded RV curve of the Keplerian signal. \textbf{Middle panels: }Residuals after fitting and their periodogram.      \textbf{Bottom panel:} Zoom on the RV measurements of the different high cadence campaigns along with the best model fit from the joint GP mode of the FWHM and the RVs.}
        \label{rv_full_dataset}
\end{figure*}

\begin{figure}
        \includegraphics[width=9cm]{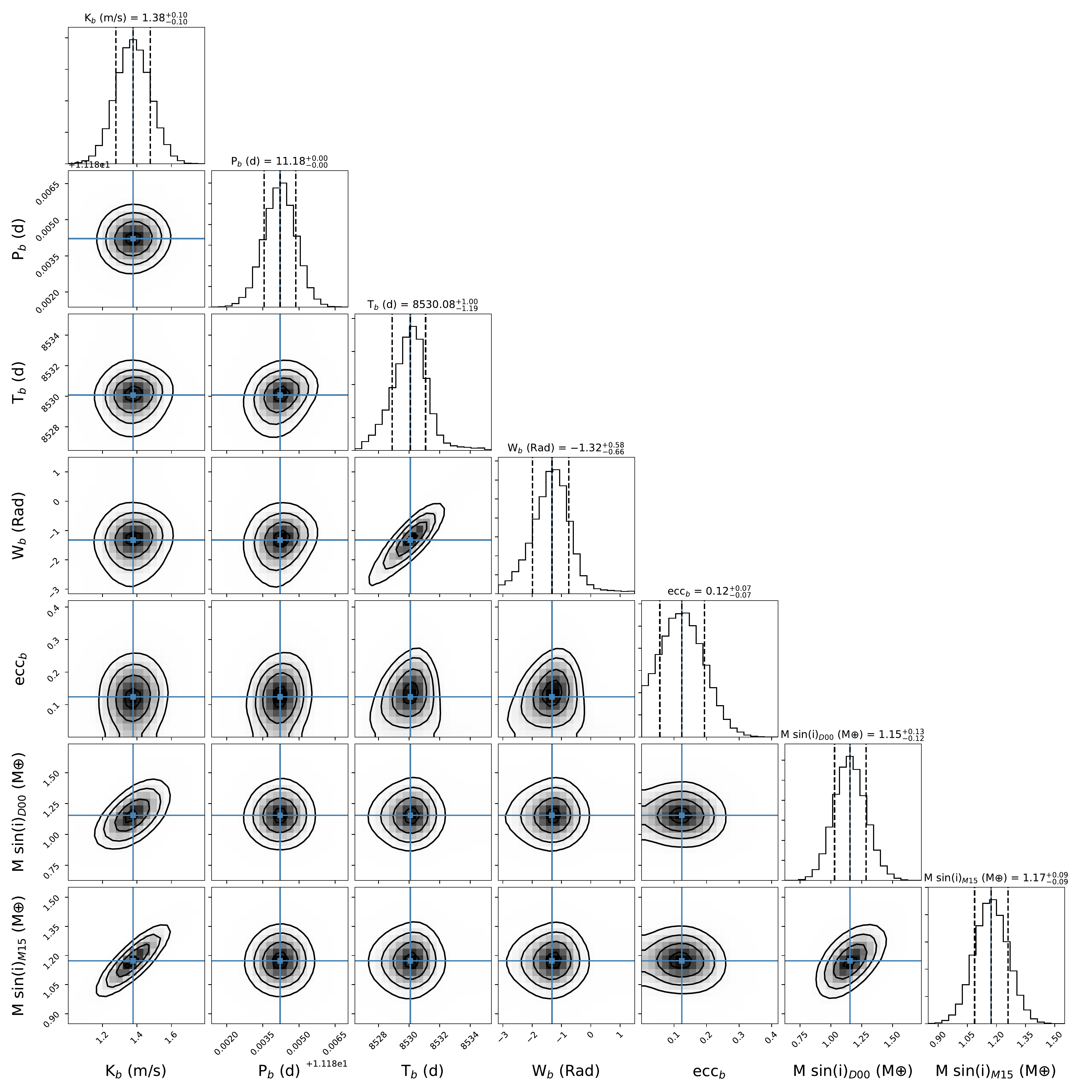}
        \caption{Posterior distributions of the parameters of the Keplerian signal at 11.18 days, from the joint model of the FWHM and the RV data, using the complete RV dataset. Figure~\ref{post_dist_1p_full} in Appendix~\ref{ap:post_dist} shows the full posterior distribution for all parameters.}
        \label{post_dist_kep}
\end{figure}

We adopted the 1-planet model with the GP as the best model to fit our RV data. Figure~\ref{rv_full_dataset} shows the model of the full dataset and the phase folded plot of the RV variations of Proxima b. Table~\ref{tab:1p_parameters} shows the final parameters obtained and the priors used in the MCMC simulations and Fig.~\ref{post_dist_kep} shows the posterior distribution of the planetary parameters. The inclusion of the HARPS and UVES archival data strengthens the significance of the detection of Proxima b, as expected, and tightens the confidence intervals of the parameters. 

\begin{table*}
\begin{center}
\caption{Parameters of the model using the full dataset. Mass $\cdot$ sin(i)$_{D00}$ shows the minimum mass of Proxima b using the \citet{Delfosse2000} estimation for the mass of Proxima, and  Mass $\cdot$ sin(i)$_{M15}$ shows the minimum mass using the \citet{Mann2015} estimation.  \label{tab:1p_parameters}}
\begin{tabular}{ l c c c c}
\textbf{Parameter} & \textbf{Priors} & \textbf{GP} & \textbf{GP + 1 Signal}& \textbf{GP + 2 Signals}\\
\textit{Keplerian} \\
Period [days] &  
$\mathcal{U}$ (2 , 20) &  & 11.18427$^{+0.00066}_{-0.00070}$ & 11.18418$^{+0.00068}_{-0.00074}$  \\
K [m$\cdot$s$^{-1}$] &  $\mathcal{U}$ (0 , 5) & & 1.377$^{+0.100}_{-0.099}$ & 1.368$^{+0.099}_{-0.109}$\\
e & $\mathcal{U}$ (0 , 1) & & 0.124$^{+0.070}_{-0.068}$ & 0.109$^{+0.076}_{-0.068}$\\
$T_{P} - 2450000$ [d] & $\mathcal{U}$ (8525 , 8535) & & 8530.1 $^{+1.0}_{-1.2}$ & 8530.2 $^{+1.3}_{-1.4}$ \\
$\omega$ [rad] & $\mathcal{U}$ (-$\pi$ , $\pi$)& & -- 1.32 $^{+0.57}_{-0.66}$   & -- 1.19 $^{+0.73}_{-0.74}$\\ \\
\textit{Derived parameters} \\
Semi-major axis [AU] & & & 0.04855 $^{+0.00029}_{-0.00029}$ & 0.04857 $^{+0.00029}_{-0.00029}$ \\
Mass $\cdot$ sin(i)$_{D00}$ [M$_{\oplus}$] &&& 1.15 $^{+0.13}_{-0.12}$ & 1.16 $^{+0.13}_{-0.13}$ \\
Mass $\cdot$ sin(i)$_{M15}$ [M$_{\oplus}$] &&& 1.173 $^{+0.086}_{-0.086}$ & 1.173 $^{+0.087}_{-0.090}$ \\
\\
\textit{Sinusoidal} \\
Period [days] &  
$\mathcal{U}$ (2 , 20) &  & &5.168 $^{+0.051}_{-0.069}$\\
K [m$\cdot$s$^{-1}$] &  $\mathcal{U}$ (0 , 5) & & & 0.35 $^{+0.10}_{-0.11}$\\
$T_{0} - 2450000$ [d] & $\mathcal{U}$ (8525 , 8535) & & & 8530.47$^{+1.25}_{-0.47}$ \\
 \\
\textit{Activity model} \\
B$_{GP}$ RV [(ms$^{-1})^{2}$] & $\mathcal{LU}$ (0.1 , 50) &  3.13 $^{+1.09}_{-0.83}$ &  4.63 $^{+1.21}_{-0.95}$&  4.65 $^{+1.24}_{-0.98}$\\
B$_{GP}$ FWHM [(ms$^{-1})^{2}$]& $\mathcal{LU}$ (0.1 , 300)&  191 $^{+50}_{-39}$&  152 $^{+32}_{-25}$ &  153 $^{+33}_{-24}$\\
P$_{GP}$ [d] & $\mathcal{LU}$ (30 , 300) & 87.0 $^{+9.6}_{-4.6}$& 87.5 $^{+12.8}_{-8.9}$ & 87.0 $^{+12.3}_{-9.4}$  \\
L$_{GP}$ [d]& $\mathcal{LU}$ (30 , 300) & 109 $^{+46}_{-33}$ & 67 $^{+20}_{-14}$ & 68 $^{+22}_{-14}$     \\
log C$_{GP}$ & $\mathcal{U}$ (-600 , 600) &  -- 300$^{+200}_{-200}$   &  -- 300$^{+205}_{-200}$  &  -- 300$^{+210}_{-200}$ \\ \\
\textit{Noise} \\
Jit$_{UVES}$ RV [m$\cdot$s$^{-1}$]& $\mathcal{LU}$ (0 , 5) & 1.47 $^{+0.22}_{-0.21}$ &  0.72 $^{+0.27}_{-0.25}$ &  0.72 $^{+0.28}_{-0.25}$ \\
Jit$_{H~pre15}$ RV [m$\cdot$s$^{-1}$]& $\mathcal{LU}$ (0 , 5) & 1.41 $^{+0.25}_{-0.25}$ &  0.78 $^{+0.29}_{-0.29}$&  0.76 $^{+0.31}_{-0.29}$ \\
Jit$_{H~post15}$ RV [m$\cdot$s$^{-1}$] & $\mathcal{LU}$ (0 , 5) & 0.92 $^{+0.23}_{-0.24}$ &  0.108 $^{+0.336}_{-0.087}$  &  0.093 $^{+0.27}_{-0.074}$\\
Jit$_{E~pre}$ RV [m$\cdot$s$^{-1}$]& $\mathcal{LU}$ (0 , 5) & 1.15$^{+0.18}_{-0.15}$ & 0.33 $^{+0.15}_{-0.14}$ & 0.213$^{+0.145}_{-0.085}$ \\ 
Jit$_{E~post}$ RV [m$\cdot$s$^{-1}$]& $\mathcal{LU}$ (0 , 5) & 1.22 $^{+0.36}_{-0.28}$ & 0.058$^{+0.16}_{-0.041}$ &0.065$^{+0.220}_{-0.046}$ \\ \\
Jit$_{H~pre15}$ FWHM [m$\cdot$s$^{-1}$]& $\mathcal{LU}$ (0 , 5) & 2.11$^{+0.59}_{-0.58}$ & 2.03$^{+0.61}_{-0.64}$ & 2.02$^{+0.61}_{-0.65}$\\
Jit$_{H~post15}$ FWHM [m$\cdot$s$^{-1}$]& $\mathcal{LU}$ (0 , 5) & 2.21 $^{+0.55}_{-0.59}$ & 1.96 $^{+0.58}_{-0.69}$& 1.98 $^{+0.56}_{-0.66}$\\
Jit$_{E~pre}$ FWHM [m$\cdot$s$^{-1}$]& $\mathcal{LU}$ (0 , 5) & 0.071 $^{+0.212}_{-0.053}$ & 0.070$^{+0.205}_{-0.051}$ & 0.064$^{+0.219}_{-0.045}$\\ 
Jit$_{E~post}$ FWHM [m$\cdot$s$^{-1}$]& $\mathcal{LU}$ (0 , 5) & 0.101 $^{+0.448}_{-0.080}$ & 0.105$^{+0.472}_{-0.083}$  & 0.113$^{+0.476}_{-0.092}$\\ \\
\textit{Polynomial terms} \\
V0$_{UVES}$ RV [m$\cdot$s$^{-1}$]& $\mathcal{U}$ (-10 , 10) & 0.1 $^{+1.4}_{-1.4}$ & 0.6$^{+1.5}_{-1.4}$ & 0.6$^{+1.5}_{-1.4}$  \\
V0$_{H~pre15}$ RV [m$\cdot$s$^{-1}$]& $\mathcal{U}$ (-10 , 10) & 0.72 $^{+0.67}_{-0.67}$ & 0.91 $^{+0.72}_{-0.72}$ & 0.87 $^{+0.76}_{-0.68}$\\
V0$_{H~post15}$ RV [m$\cdot$s$^{-1}$]& $\mathcal{U}$ (-10 , 10)& --0.21 $^{+1.0}_{-1.0}$& -- 0.5 $^{+1.1}_{-1.1}$ & -- 0.5 $^{+1.1}_{-1.2}$  \\
V0$_{E~pre}$ RV [m$\cdot$s$^{-1}$]& $\mathcal{U}$ (-10 , 10) & -- 0.5 $^{+1.5}_{-1.5}$ & -- 1.1 $^{+1.5}_{-1.6}$ & -- 1.1 $^{+1.5}_{-1.5}$ \\ 
V0$_{E~post}$ RV [m$\cdot$s$^{-1}$]& $\mathcal{U}$ (-10 , 10) & 1.5 $^{+1.5}_{-1.5}$ & 0.8 $^{+1.7}_{-1.7}$& 0.9 $^{+1.7}_{-1.8}$\\ 
Lin  RV [m$\cdot$s$^{-1}$ $^{-d}$]& $\mathcal{U}$ (-0.1 , 0.1) & -- 0.00008 $^{+0.00039}_{-0.00039}$ & 0.00021 $^{+0.00043}_{-0.00039}$ & 0.00009 $^{+0.00041}_{-0.00043}$  \\ \\
V0$_{H~pre15}$ FWHM [m$\cdot$s$^{-1}$]& $\mathcal{U}$ (-10 , 10) & 5.7 $^{+5.2}_{-5.2}$ & 6.1 $^{+4.4}_{-4.4}$ & 5.9 $^{+4.4}_{-4.2}$\\
V0$_{H~post15}$ FWHM [m$\cdot$s$^{-1}$]& $\mathcal{U}$ (-10 , 10) & --3.6 $^{+8.1}_{-7.9}$ & --4.1 $^{+6.7}_{-6.7}$& --3.9 $^{+6.7}_{-6.8}$\\
V0$_{E~pre}$ FWHM [m$\cdot$s$^{-1}$]& $\mathcal{U}$ (-10 , 10) & -- 4 $^{+11}_{-11}$ & -- 4.9 $^{+9.4}_{-9.5}$ & -- 4.9 $^{+9.8}_{-9.1}$\\ 
V0$_{E~post}$ FWHM [m$\cdot$s$^{-1}$]& $\mathcal{U}$ (-10 , 10) & -- 8 $^{+12}_{-12}$  & --9 $^{+10}_{-10}$  & --10 $^{+11}_{-10}$ \\ 
Lin  FWHM [m$\cdot$s$^{-1}$ $^{-d}$]& $\mathcal{U}$ (-0.1 , 0.1) & 0.0033 $^{+0.0031}_{-0.0031}$ & -- 0.0035 $^{+0.0027}_{-0.0027}$ & -- 0.0035 $^{+0.0027}_{-0.0027}$\\ \\
\textit{Residuals} \\
RMS res [m$\cdot$s$^{-1}$] && 1.57 & 1.10  & 1.07 \\
RMS res$_{UVES}$ [m$\cdot$s$^{-1}$] && 1.48 & 0.78  & 0.79 \\
RMS res$_{HARPS}$  [m$\cdot$s$^{-1}$] && 1.72 & 1.33  & 1.30 \\
RMS res$_{ESPRESSO}$  [m$\cdot$s$^{-1}$] && 1.15 & 0.37  & 0.28 \\\\
$log~Z$ & & -- 1469 & -- 1405 & -- 1398 \\\hline
\end{tabular}
\end{center}
\end{table*}

We measure a minimum mass of  1.173 $\pm$ 0.086 M$_{\oplus}$, compatible with the measurements of \citet{AngladaEscude2016}, although slightly smaller and with slightly smaller uncertainties. On the other hand, we measure a slightly larger mass than the measurement of \citet{Damasso2020}, but it could be caused by the different choice of model. We only consider a single planet, described by a Keplerian, while \citet{Damasso2020} consider the presence of two planets, described as sinusoidals, in the system. The GP kernel in both cases is also different, potentially causing small differences between the different measurements. Our amplitude and period measurements perfectly coincide with the original measurements, while our eccentricity limit is smaller. The combined dataset constrains the eccentricity to be smaller than 0.19. 

The GP in this case converges to a rotation period of 87 $\pm$ 12 days, with a decay timescale of $\sim$2/3 of a rotation. This result is compatible with our expectations and with previous measurements of both the rotation period and the decay timescale~\citep{Masca2016, Giles2017}. 

\subsubsection{Comparing ESPRESSO and HARPS}

The similar design philosophies of ESPRESSO and HARPS allow for an approximate direct comparison between the two instruments. However, comparing their performance at detecting periodic signals is not an easy task. The root mean square (RMS) of the residuals would suggest that ESPRESSO is performing 3.6 times better than HARPS but that might not tell the whole story. Following \citet{HatzesCochran1992 and  Figueira2018} we perform a back-of-the-envelope calculation. Assuming that the noise on the spectrum is photon noise only, and that the spectrum is characterised by a uniform density of lines, the authors concluded that the RV error is given by

\begin{equation}
\sigma_{RV} \propto {1 \over {\sqrt{F} \sqrt{\Delta \lambda} R^{1.5}}}
,\end{equation}

\noindent where F is the average flux level, $\Delta$ $\lambda$ is the wavelength coverage, and R is the
resolution. Comparing the ESPRESSO and HARPS characteristics we find a resolution factor of (140/115)$^{1.5}$ = 1.34, a flux factor of 2.30 (coming from the telescope size), and a wavelength coverage factor of 1.14. This leads to an improvement in precision of a factor of 3.51, very close to the measured difference in scatter of the residuals. It is important to restate that this is valid only for the photon-noise-limited regime.

\begin{figure}
        \includegraphics[width=9.0cm]{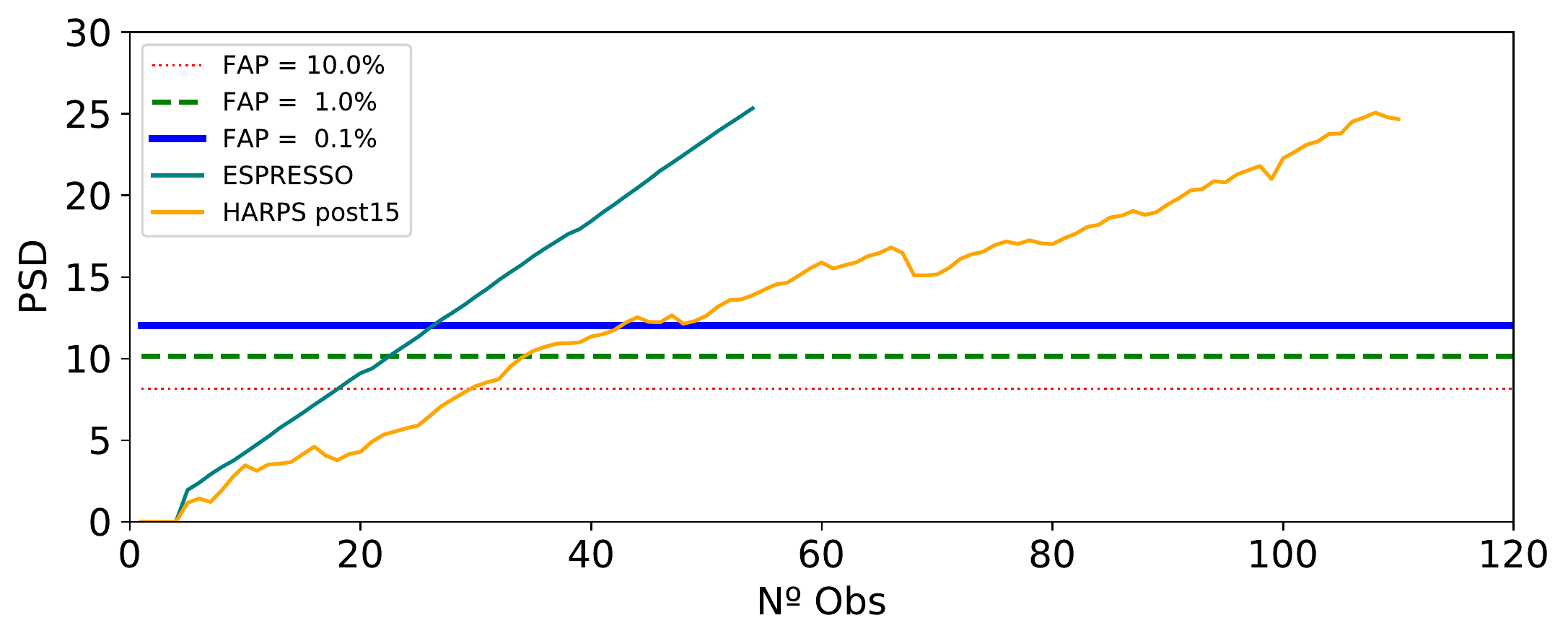}
        \caption{Evolution of the periodogram power for the 11.2 days signal in ESPRESSO and the HARPS RedDots campaigns.}
        \label{perf_vs_harps}
\end{figure}

Again using a stacked periodogram \citep{Mortier2017} we compute the evolution of the periodogram power as we include new measurements for ESPRESSO and HARPS independently. We use the HARPS 2016 and 2017 campaigns, as their observational strategy is the closest to our own, although we must take into account that Proxima's induced stellar RV signal was simpler in shape (easier to model) during those two campaigns than it was in 2019 (see Sect. ~\ref{sect:activity}). The slope at which the PSD increases gives us a rough idea of the performance of the instrument at extracting information about the RV variations. As before, we model the activity and the Keplerian simultaneously, and subtract the activity component. Figure ~\ref{perf_vs_harps} shows the results of the evolution of both datasets. As the significance of the detection and the PSD are not exactly the same thing, the HARPS PSD values have been scaled so it shares the same FAP levels as the ESPRESSO PSD. We see that the slope obtained in the ESPRESSO dataset is at least two times steeper, and much steadier,  than in the HARPS dataset. 

This is by no means a definitive assessment of the performance difference between the instruments, as it relies on the quality of the activity model in order to be meaningful. Still it could serve as a first order comparison with a real case. We find that for the case of a mid-M dwarf, ESPRESSO in its current state is capable of detecting planetary signals with at least two times less data than HARPS. This number should increase as we move to lower amplitude RV signals, and closer to the limit of HARPS, until we move into a regime where the detection is impossible with HARPS but possible with ESPRESSO. 

The different jitter measurements we obtain in our global measurement also tell us that, for the case of Proxima, HARPS was not limited by the 3.6m ESO Telescope, but by its own stability limit when observing without simultaneous calibration. Even at ~1-2 m$\cdot$s$^{-1}$ photon noise, an extra 1 m$\cdot$s$^{-1}$ jitter is required to fit the data. In ESPRESSO at 0.26 m$\cdot$s$^{-1}$ photon noise we obtain a jitter measurement compatible with zero, suggesting the data is truly photon noise limited. 

We must again stress that it is hard to extrapolate this behaviour to a general case. We are working in a regime where ESPRESSO seems photon noise limited, while in the case of HARPS this limitation might be exaggerated by the lack of simultaneous calibration. While Proxima's stellar activity is less problematic than its reputation suggests, it still creates a large amplitude signal that is the main cause of RV variability. Brighter and quieter stars would be more appropriate to properly compare the performance of HARPS and ESPRESSO.

\section{Stellar activity}  \label{sect:activity}

The stellar activity of Proxima has been widely studied during recent years. The star is known to have a mean rotation period of the order of 83 days \citep{Benedict1998,Masca2016}, showing evidence of differential rotation from 75-95 days  \citep{Wargelin2017}. There is also evidence for the presence of a seven-year cycle in photometry and X-rays that is assumed to be solar-like \citep{Masca2016, Wargelin2017}. 

In addition to the spectroscopic measurements obtained from the same data as the RV, we accumulated 120 days of V-band measurements using the LCO network, which we can add to the original LCO data obtained during the RedDots campaign. While the baseline is rather short, it allows us to independently recover a periodicity of 85 days, similar to previous estimations of the rotation period and to our own estimation combining RV and FWHM. We measure a peak-to-peak change in flux larger than 10\% of the V-flux. If we assume a $\Delta$T of 300K, as found in the Doppler imaging study of \citet{Barnes2015}, it would correspond to a $\Delta$ filling factor larger than 50\% of the stellar disc. This $\Delta$ filling factor, while large, is compatible with filling factor measurements obtained using Doppler tomography of rapidly rotating M-dwarfs \citep{Barnes2004}.

\begin{figure}
        \includegraphics[width=9.0cm]{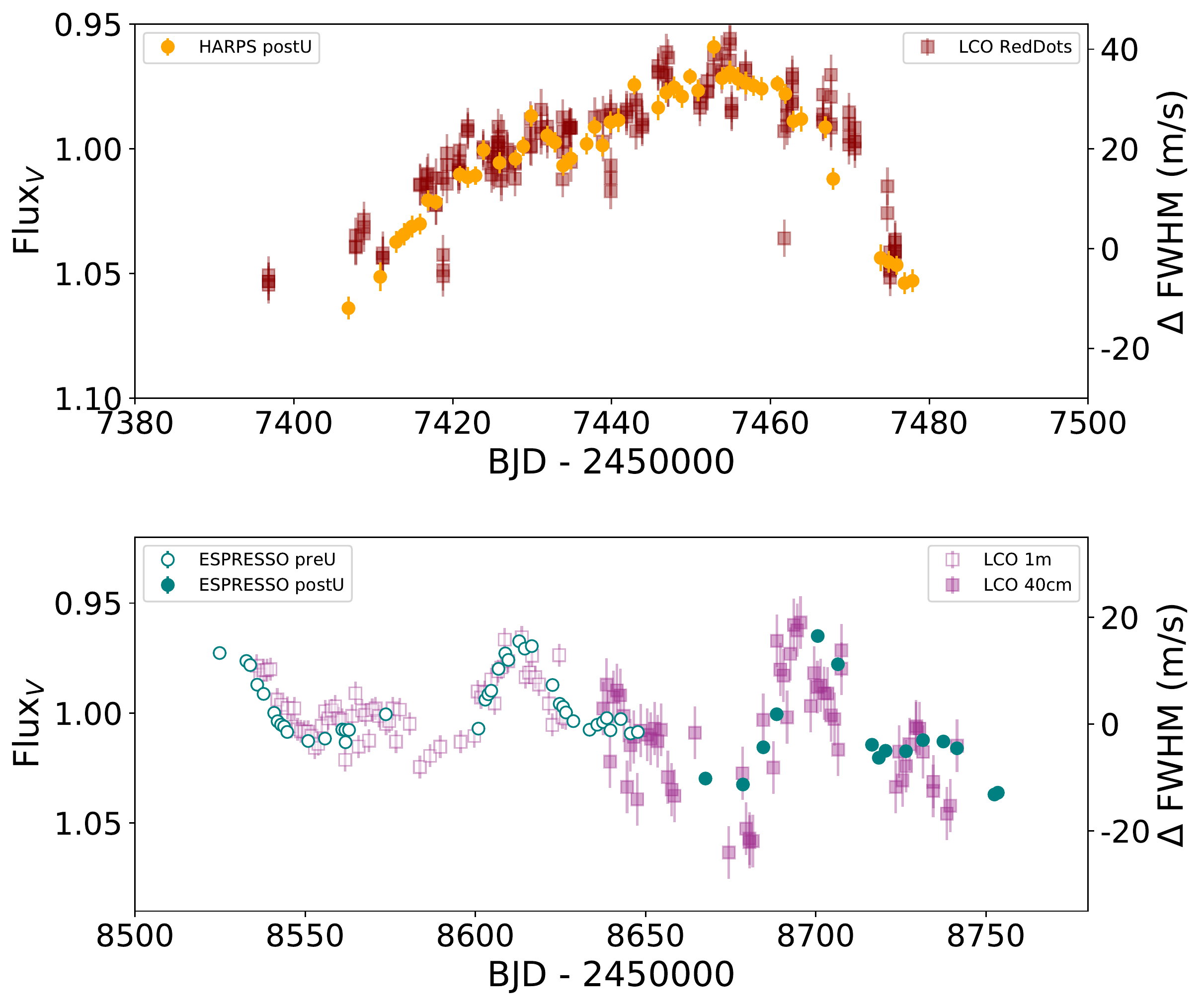}
        \caption{V-band flux and FWHM of the CCF superimposed for the 2016 RedDots campaign (top panel) and the 2019 ESPRESSO campaign (bottom panel).}
        \label{fwhm_lco}
\end{figure}

The variations measured in the FWHM of the CCF of ESPRESSO and HARPS show a very similar behaviour to the LCO photometry. Figure~\ref{fwhm_lco} shows the FWHM and the V-band LCO measurements superimposed. The FWHM in this case is acting as pseudo-photometry, with  precision as good the LCO photometry itself, and simultaneous to the RV data. This makes it the preferred activity indicator for our global model. 

The global model described in Sect.~\ref{sec:prox_b} converged to a rotation period of 87 days with a decay timescale of 66 days. The variations in the FWHM showed an amplitude of 12 m$\cdot$s$^{-1}$ . The rotation period measurement is consistent with previous measurements, while the decay timescale is also consistent with what would be expected for a main sequence star \citep{Giles2017}.

\begin{figure}
        \includegraphics[width=9.0cm]{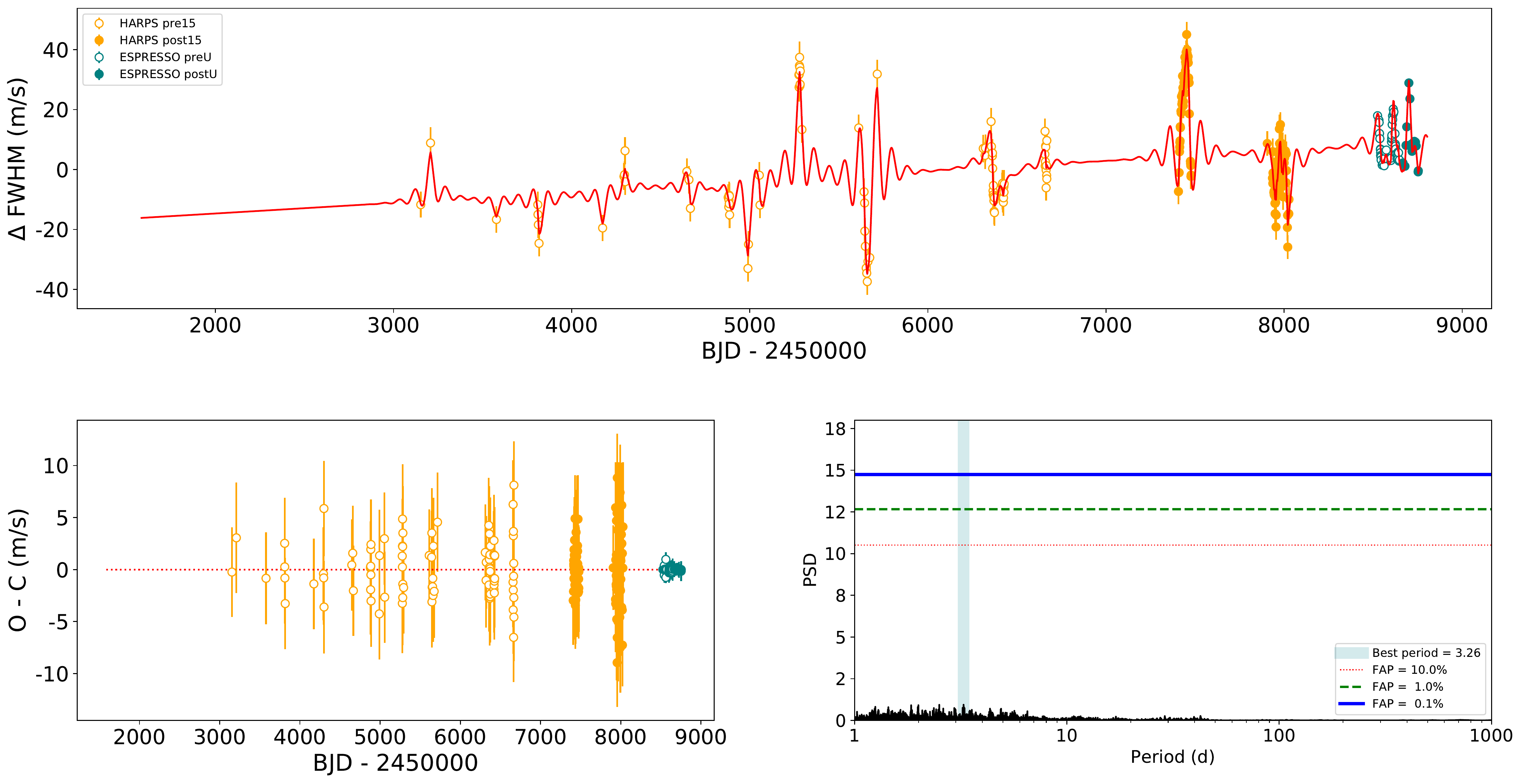}
\caption{\textbf{Top panels:} Full FWHM dataset along with the best fit using a GP activity model. \textbf{Middle panels: }Residuals after fitting and their periodogram.    \textbf{Bottom panel:} Zoom on the RV measurements of the different high cadence campaigns along with the best model fit from the joint GP mode of the FWHM and the RVs.}
        \label{fwhm_activity}
\end{figure}

\begin{figure*}
        \includegraphics[width=18.0cm]{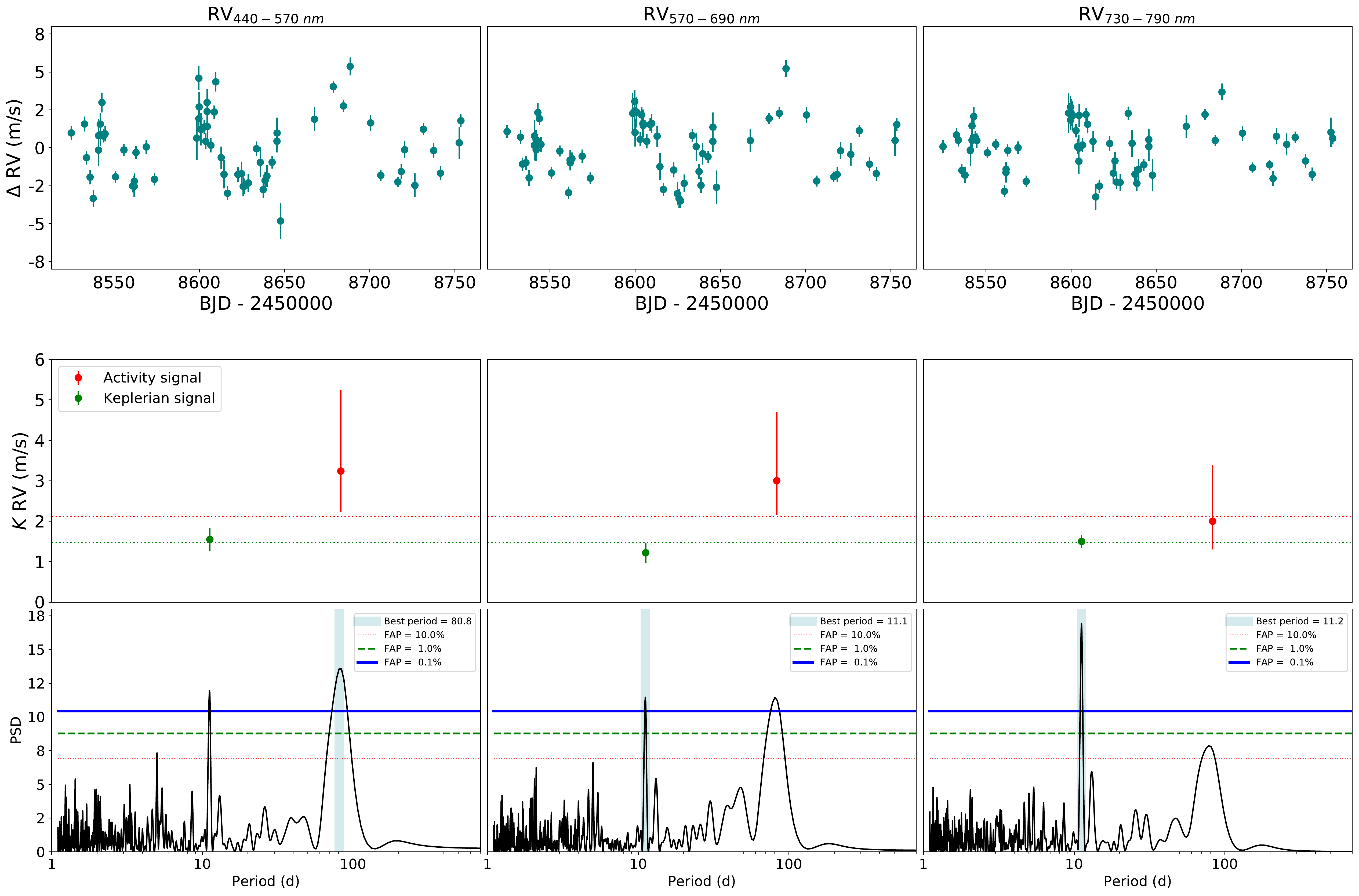}
        \caption{\textbf{Top panels:} RV data measured at different wavelength bins. \textbf{Middle panels:} Amplitude of the Keplerian and activity signals for the different wavelength bins. The horizontal lines show the measurement using the complete wavelength range. \textbf{Bottom panels:} GLS periodograms of the data at the same wavelength bins.}
        \label{croma_data}
\end{figure*}

The different observation campaigns also show wildly different shapes of the variation, suggesting strong variations in the geometry of the active regions on Proxima's surface. While the spot filling factor remains similar, the spot configuration seems very different. The 2016 RedDots campaign shows a smooth fainting over most of the campaign, with the flux recovering at the end. During the 2019 campaign the flux changed more abruptly, producing an almost triangular shape in the light curve. The measurements of the 2016 campaign suggest a surface dominated by large spots on one face of the star, with the spot coverage shrinking to the other face. The shape of the light curve in the 2019 campaign on the other hand suggests a very abrupt imbalance in the spot coverage between the two faces. While some gaps in the coverage of the full rotation, and reduced photometric quality once moved to the 0.4 m LCO telescopes, do not allow for a definitive statement, there is a hint of the star showing the "double-dipping" behaviour described in \citet{Basri2018}. This behaviour could be explained by either the presence of short-lived spots or long-lived spots, and a change in the spot configuration caused by differential rotation. A longer baseline of high quality photometry would be needed to confirm this. The abrupt changes in the light curve (and the FWHM of the CCF) during the 2019 campaign might also explain why the GP model in Sect.~\ref{sec:prox_b_esp} converged to 47 days instead to the true rotation period of the star, as it did using the full dataset.

Contrary to what has been found in long-term photometric time series, the long term FWHM time series does not show clear evidence of a magnetic cycle. This comes as a bit of a surprise, since the FWHM tracks the photometry very closely along individual rotations, and there is a visual indication of a long period variation just by looking at the data. A possible explanation is that any possible long period variation gets suppressed by the floating means of the different datasets. None of the datasets share common observing runs, making the determination of the relative zero points (both in FWHM and RV) model dependent. The only individual dataset long enough to show a full phase of the cycle on its own is the HARPS pre2015 campaign, which is also the more sparsely sampled. While there seems to be a slope between the first and second HARPS post-2015 campaigns, it is not enough to link it with the presence of a long period cycle.

While the very non-sinusoidal shape of the rotation signal can create false positives at its harmonics and their aliases, we find no indication of activity changes at timescales comparable with the period of Proxima b, neither in the FWHM nor in the photometric data. We believe this to be the reason why the GP converges to periods around 45 days when using only ESPRESSO data. Earlier this year Proxima was observed by the Transiting Exoplanet Survey Satellite (TESS) as part of Sectors 11 and 12. We did find evidence of a slope compatible with the rotation, but did not find evidence for a photometric variation at timescales of 10-12 days. \citet{Vida2019} analysed the data, finding very similar results. 

\subsection{Chromatic radial velocity variations}

The mask used to produce the RVs is comprised of thousands of spectral lines distributed between 440 - 790 nm wavelengths. Given the wide range of ESPRESSO, reaching 100 nm redder than HARPS, and the great collecting power of the VLT, we can split the spectra into smaller wavelength bins to produce different velocity measurements. We create three independent RV series using the ranges of 440-570 nm, 570-690 nm, and 730-790 nm. We define them this way because they retain a similar RV noise level, of the order of 0.5 m$\cdot$ s$^{-1}$ , for the three time series. The data bluer than 440 nm are not used for M-dwarfs due to the low flux measured, and the section between 690-730 nm is not used because of telluric contamination. 

Figure~\ref{croma_data} shows the velocities measured in the different wavelength ranges and the periodogram for each series. We find that the red RV series shows a much smaller RMS than the other two bands, and smaller than the global RV data measured using the complete spectral range. From bluer to redder, we measure scatter of the RVs of 2.35, 2.10, and 1.67 m$\cdot$ s$^{-1}$. We also see that the balance between the rotation-induced signal and the Keplerian signal changes when moving to the redder velocities. The two blue series show very similar results, with the $\sim$80d signal showing similar significance to the $\sim$11d signal. In the red time series the situation is the opposite, the $\sim$11d becomes much more significant than the $\sim$80d signal. The 5.15 days signal we find when modelling the ESPRESSO data  might also show a chromatic evolution, but its significance is too low in the individual datasets to make any definitive claim.

\begin{figure}
        \includegraphics[width=9.0cm]{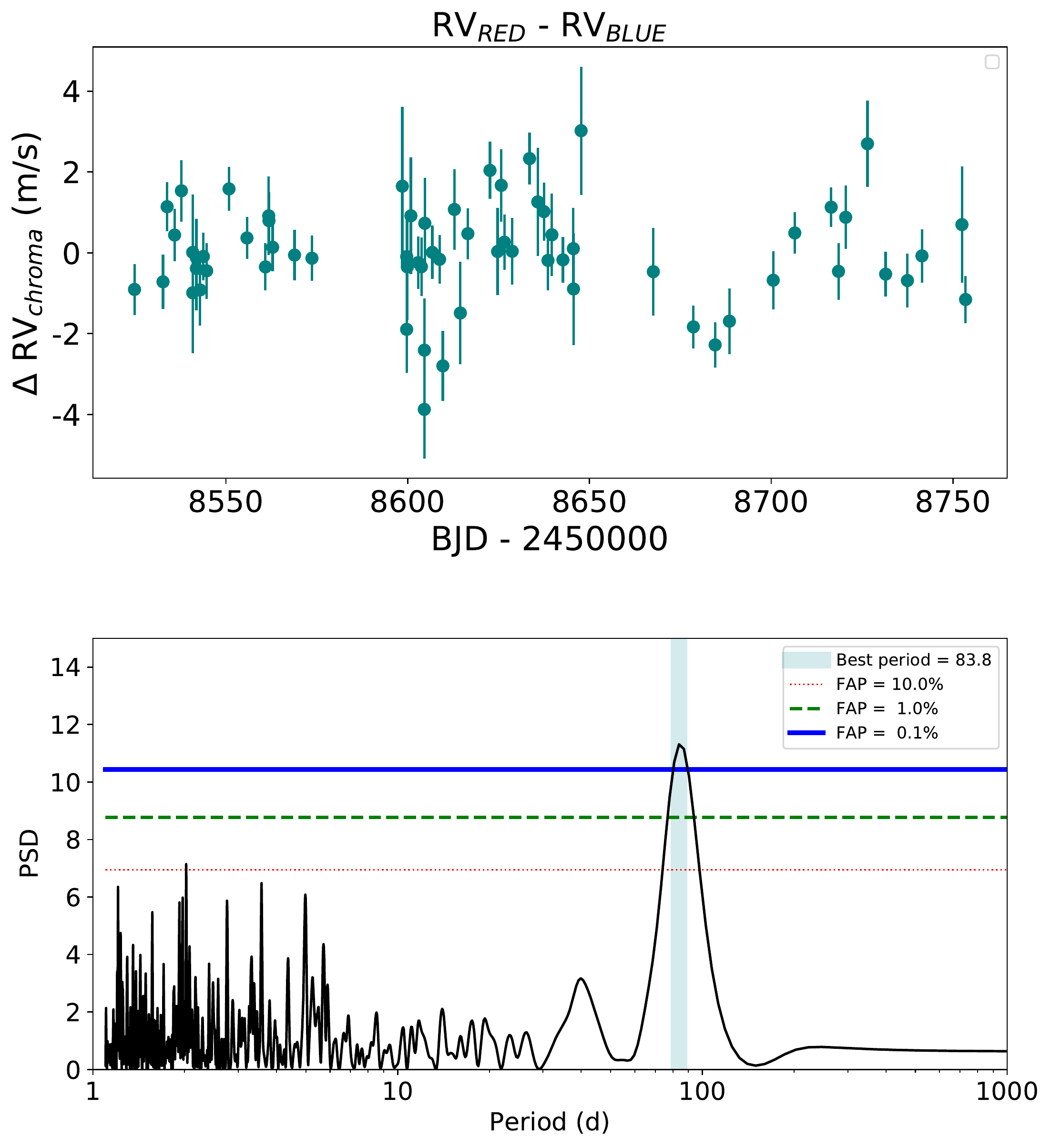}
        \caption{Chromatic RV differences and their periodogram.}
        \label{croma_drv}
\end{figure}

The difference in balance between the two signals in the different spectral ranges suggests the RV information we obtain is different at different wavelengths. This is not unexpected, as spot-induced RV variations are known to diminish at redder wavelengths and this effect has already been measured for very active stars \citep{Figueira2010, Zechmeister2018}. We analyse a \textit{chromatic} RV by measuring the difference between the reddest RV (730-790 nm) and the bluest RV (440-570 nm). Our chromatic RV shows a scatter of 2 m$\cdot$ s$^{-1}$, larger than the photon noise of the measurements (assumed as the quadratic sum of the red photon noise and the blue photon noise). A GLS periodogram shows a clear peak at $\sim$83 days, very close to the measured rotation period of the star, while no indication of any signal at the $\sim$11 days range. Figure~\ref{croma_drv} shows the chromatic RV and its GLS periodogram. The suppression of the 11d signal indicates the RV information related to this signal is the same in the two wavelength ranges, as expected for a Keplerian signal. Conversely the RV information related to the 80d signal seems to have a chromatic component. This is further evidence that it is purely an activity signal and, combined with its diminished strength in the red wavelengths, that it is spot induced. 

The extended wavelength range of ESPRESSO, when compared to HARPS, and the increased collecting power of the VLT open the possibility of producing RVs at different wavelength ranges without sacrificing too much photon noise precision. This makes it possible to study the chromatic RV differences related to activity-induced signals even at amplitudes lower than 3 m$\cdot$ s$^{-1}$ like we have just demonstrated. While it goes beyond the scope of this article, these capabilities open the possibility of using custom spectral slices to create RV time series that minimise (or maximise) the contribution of the activity-induced signals while maintaining a good photon noise precision even for stars fainter than Proxima. A less sophisticated, but much easier to implement, version of the idea proposed by \citet{Dumusque2018,  Cretignier2019} .

\subsection{Relation between the FWHM of the CCF and RV in Proxima}

\begin{figure}
        \includegraphics[width=9.0cm]{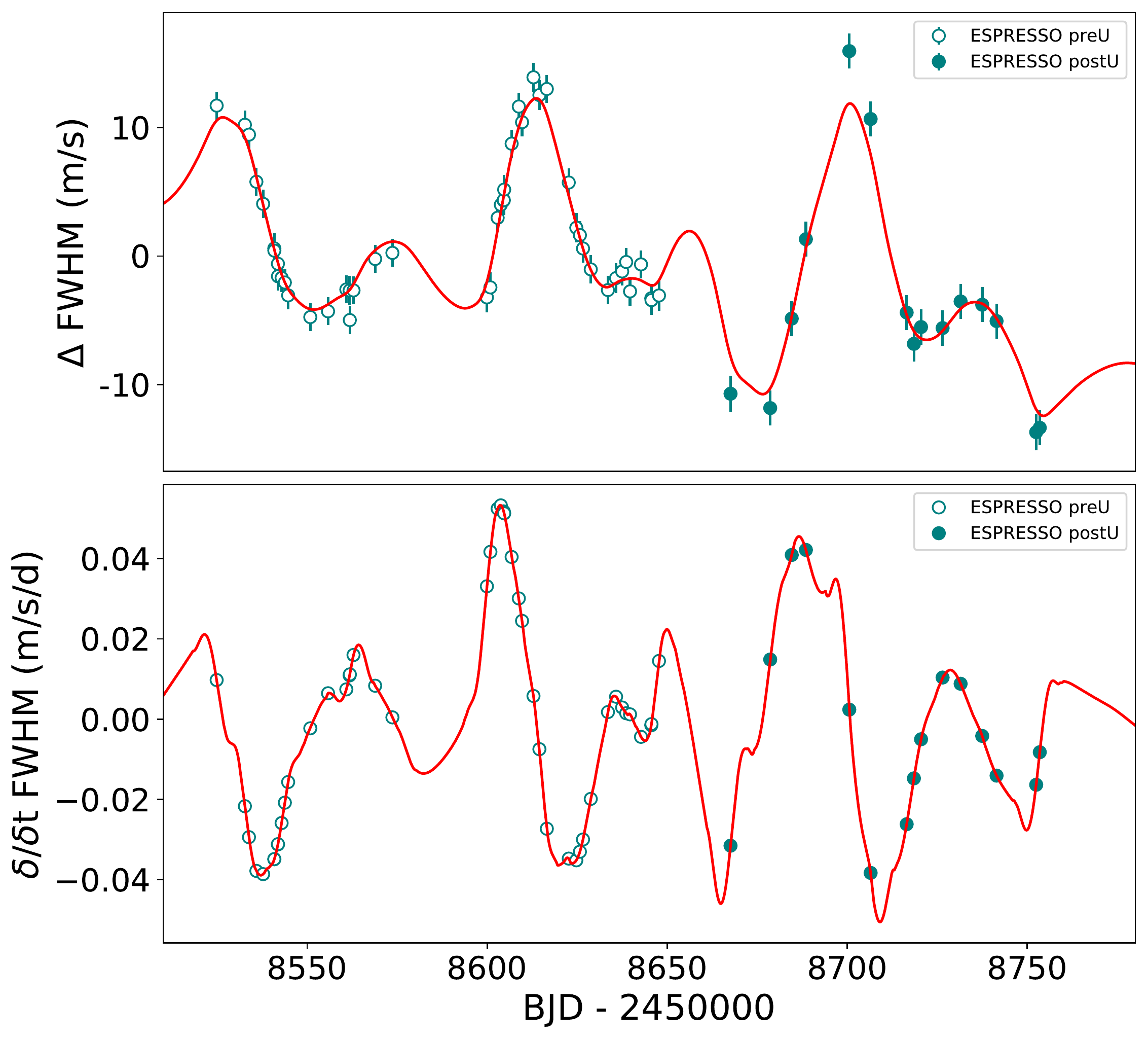}
        \caption{Full width half maximum of ESPRESSO with the best model and its derivative.}
        \label{fwhm_dev}
\end{figure}

In simple cases, the RV variations caused by stellar activity can be modelled as a combination of the stellar flux and its derivative (e.g. the F/F' method proposed by \citet{Aigrain2012}). As seen in Sect.~\ref{sect:activity}, we can use the FWHM of the CCF as a proxy for photometry, providing us with a simultaneous measurement of RV and the flux variations for all the HARPS and ESPRESSO measurements. Using the model derived previously we can calculate its derivative and test whether or not the FWHM can provide similar information. Figure~\ref{fwhm_dev} shows the ESPRESSO FWHM data and its derivative. 

\begin{figure}
        \includegraphics[width=9.0cm]{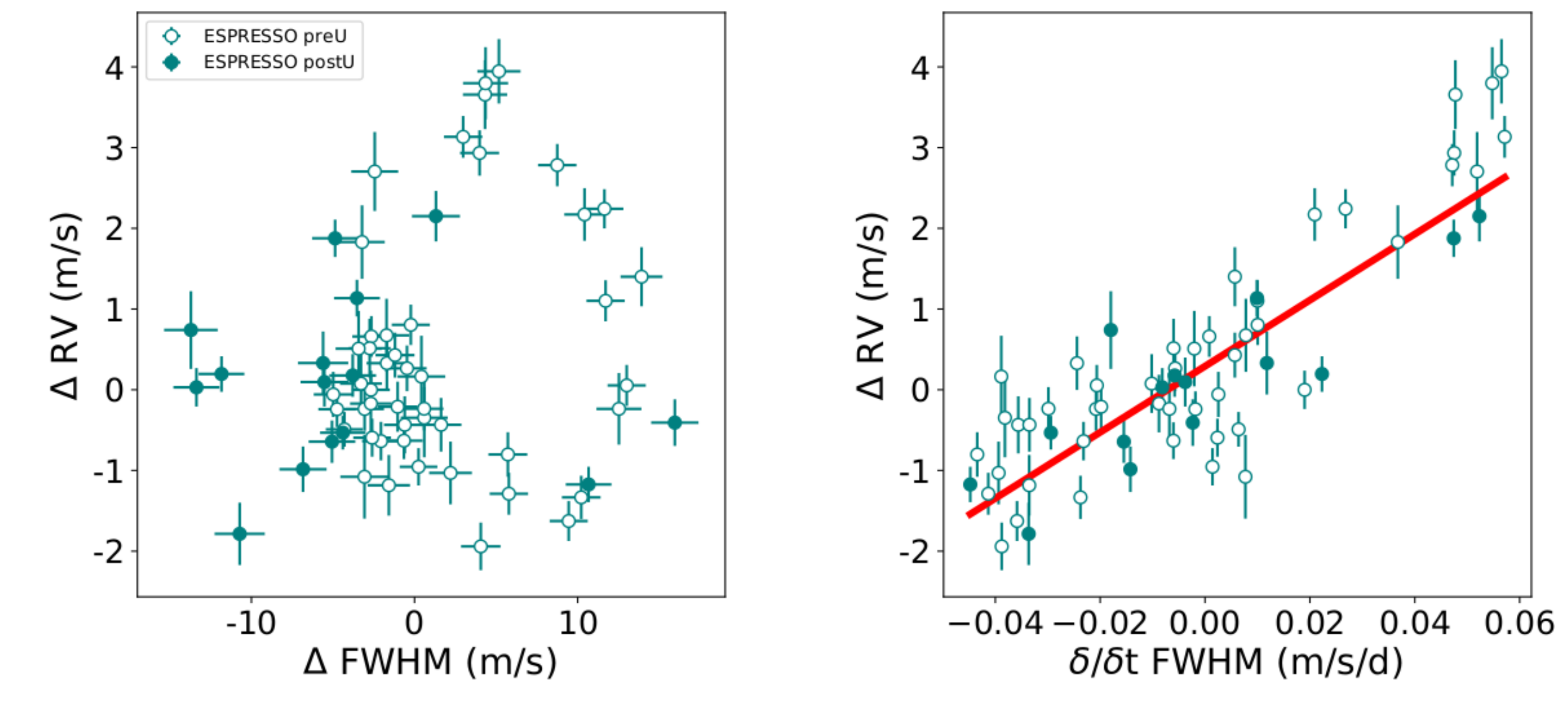}
        \caption{Relation between the activity-induced RV variations, the FWHM, and the derivatives of their respective best models. }
        \label{correl}
\end{figure}

When comparing the FWHM and its derivative against the activity component of the RV, we found the two behaved very differently. The variations of the FWHM and the activity-induced RV form a sort of circular shape, while the derivatives of the FWHM and the RV show a clear linear trend. Figure~\ref{correl} shows the different correlations found in the data.

\begin{figure}
        \includegraphics[width=9.0cm]{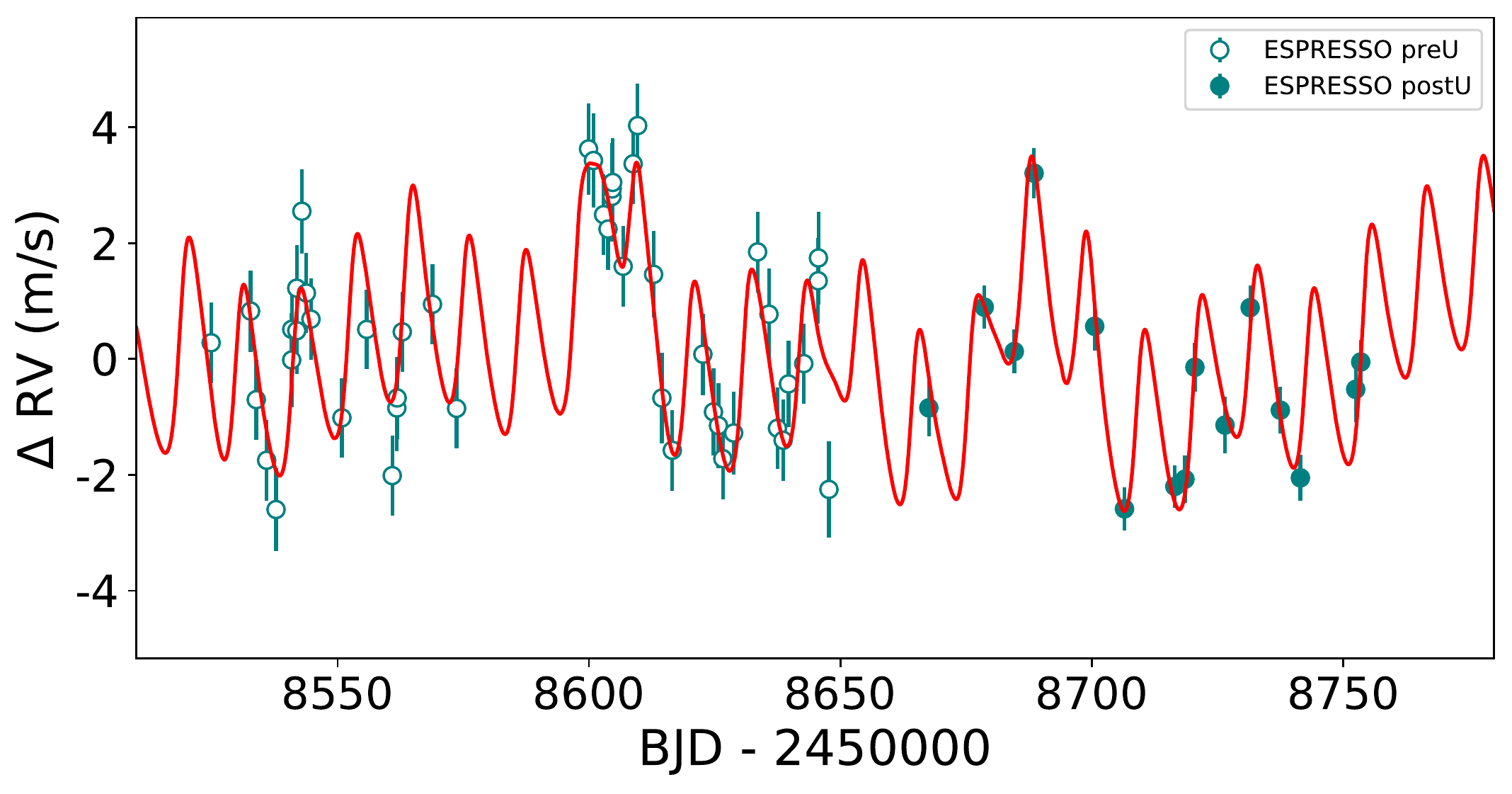}
        \caption{ESPRESSO RVs with the best model obtained using the derivative of the FWHM as activity proxy. }
        \label{rv_correl}
\end{figure}

Using the derivative of the FWHM as our activity proxy, we can fit the activity RV variations using a simple model based only on a first order polynomial against the derivative.  Figure~\ref{rv_correl} shows the model of the activity-induced RV data with the best model we obtain using the derivative of the FWHM as activity proxy with one extra Keplerian to account for the variations induced by Proxima b. We can model the RV variations with a residual of 59 cm$\cdot$s$^{-1}$. We decided not to include the candidate signal at 5.15 days, as its nature is not completely clear. This result, although not as good as the fit obtained using the GP framework, provides some extra insights into the nature of the active regions causing the RV variations. It suggests the dominant effect is a photometric effect, caused by a few large spots, or groups of spots. Otherwise a method as simple as this one would probably not work \citep{Aigrain2012}. 

\section{Are there other planets around Proxima? }

\begin{figure}
        \includegraphics[width=9.0cm]{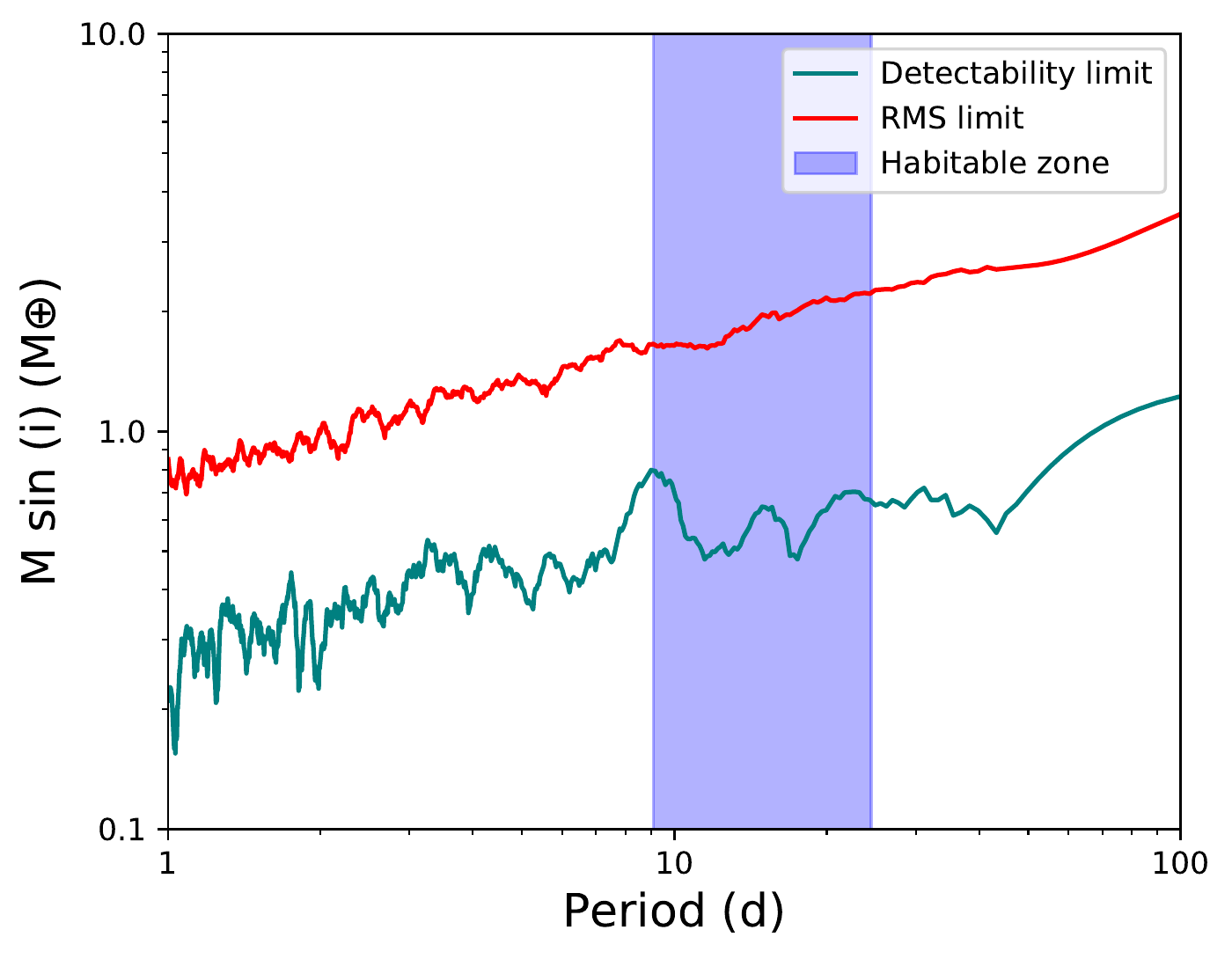}
        \caption{Detection limits for the presence of additional planets around Proxima. The red line shows the limit imposed by the total dispersion of the RV measurements. More massive planets would cause a RV dispersion too large to be explained by the data we collected.} The teal line shows the limit above which we should have detected the planets with our current dataset.  The blue region shows the periods corresponding to the habitable zone around the star. 
        \label{det_lims}
\end{figure}

The presence of additional low-mass planets in the Proxima system remains an open question. Low-mass stars tend to host multiple rocky planets, sometimes at close-in orbits. Even with the extreme quality of our measurements we could not detect any other planet in the system at periods shorter than the rotation period of Proxima. While we cannot rule out the presence of other very-low-mass planets, we can establish some quite strong constraints on the types of planets that could accompany Proxima b. The low RMS of the residuals of the ESPRESSO data do not allow for the presence of Earth-mass planets at most close orbits, and certainly not for the presence of larger planets. Figure ~\ref{det_lims} shows the results of 500~000 simulations injecting planets with orbits of 1-100 days. For each simulation we injected a sinusoidal signal in our dataset and tried to recover it. We considered a detection confirmed if we could detect it with 99\% probability in a combined model with Proxima b and the GP. 

The total dispersion of the data would not allow for the presence of any extra planets more massive than 2 M$_{\oplus}$ at the periods corresponding to the HZ or shorter, or at masses larger than 4 M$_{\oplus}$ at any orbital period shorter than 100 days. We found that we should have detected the presence of extra planets more massive than 0.4-0.5 M$_{\oplus}$ at orbital periods compatible to the HZ or shorter.

\section{Conclusions}

We revisited Proxima using the new generation spectrograph ESPRESSO at the VLT. The improved RV precision and collecting power allowed us to independently detect Proxima b with a much smaller number of observations than originally needed. Using a joint model combining activity indicators and RV data, and Gaussian processes, we confirm the presence of Proxima b at a period of 11.22 $\pm$ 0.03 days with minimum mass of 1.29 $\pm$ 0.13 M$_{\oplus}$, compatible with the result published by \citet{AngladaEscude2016}. When combining the data with the archival HARPS and UVES data we correct our measurement to a period of 11.18427$\pm$0.00066 days and a minimum mass of 1.173 $\pm$ 0.086 M$_{\oplus}$. Applying the best model to the ESPRESSO data produces an RMS of the residual of 27 cm$\cdot$s$^{-1}$ , very close to the RV photon noise level. We find evidence for a second short period signal with a period of 5.15 days and a semi-amplitude of 0.4 m$\cdot$s$^{-1}$. If caused by a planetary companion, it would correspond to a minimum mass of 0.29 $\pm$ 0.08 M$_{\oplus}$ at an orbital distance of 0.02895 $\pm$ 0.00022 AU, with an equilibrium temperature of 330 $\pm$ 30 K. Further ESPRESSO observations will be needed to confirm the presence of the signal and establish its origin. We do not detect any additional companions up to 0.4 M$_{\oplus}$ at orbital distances shorter than the HZ of the star.  

We find a rotational modulation of 87 days, compatible with previous measurements, and no clear evidence for a long period modulation present in the FWHM or the RV time series. For the case of Proxima, we find that the FWHM of the CCF of HARPS and ESPRESSO can be used as a proxy for brightness changes. It provides a precision similar to what can be obtained with 5 mmag ground-based photometry. We found that the FWHM of the CCF was able to track the brightness changes related to the rotation of Proxima with exquisite detail. The different shapes of FWHM and photometric time series obtained in different years suggest the surface of Proxima has changed a lot with time. While the filling factor seems to remain similar, the geometry of the activity regions seems to be wildly different in different years. 

The extended spectral range of ESPRESSO with respect to HARPS, combined with the collecting power of the VLT, allows us to split the spectrum into different wavelength bins to create independent RV series, while maintaining a good photon noise level in each bin. We find that we can measure the decline of a low-amplitude activity signal towards redder wavelengths, as would be expected for spot-induced variations. The planetary signal on the other hand shows a constant velocity amplitude across the full wavelength range, as is also expected for Keplerian signals. We define a chromatic RV, based on the difference between the red and blue velocities, which seems to efficiently track the activity variations of Proxima. Using the time series of the FWHM of the CCF and its gradient, we are able to model the stellar activity in a similar way to the F/F' method \citep{Aigrain2012}, obtaining good results when detrending the data from activity to recover the planetary signal. 

\begin{acknowledgements}
ASM acknowledges financial support from the Spanish Ministry of Science and Innovation (MICINN) under the 2019 Juan de la Cierva Programme. 

A.S.M., J.I.G.H., R.R., and C.A.P. acknowledge financial support from the Spanish Ministry of Science, Innovation, and Universities (MICIU) AYA2017-86389-P. 

J.I.G.H. acknowledges financial support from Spanish MICIU under the 2013 Ram\'on y Cajal programme RYC-2013-14875. 

FPE and CLO would like to acknowledge the Swiss National Science Foundation (SNSF) for supporting research with ESPRESSO through the SNSF grants nr. 140649, 152721, 166227, and 184618. The ESPRESSO Instrument Project was partially funded through SNSF’s FLARE programme for large infrastructures.

This work was financed by FEDER  (Fundo Europeu de Desenvolvimento Regional) funds through the COMPETE 2020 Operational Programme for Competitiveness and Internationalisation (POCI), and by Portuguese
funds through FCT (Funda\c c\~ao para a Ci\^encia e a Tecnologia) in the framework of the project with references POCI-01-0145-FEDER-028987 and PTDC/FIS-AST/28987/2017.

This work was supported by FCT through national funds and by FEDER through COMPETE2020 - Programa Operacional Competitividade e Internacionalização by these grants: 
UID/FIS/04434/2019; UIDB/04434/2020; UIDP/04434/2020; 
PTDC/FIS-AST/32113/2017 \& POCI-01-0145-FEDER-032113; 
PTDC/FIS-AST/28953/2017 \& POCI-01-0145-FEDER-028953; 
PTDC/FIS-AST/28987/2017 \& POCI-01-0145-FEDER-028987;
PTDC/FIS-OUT/29048/2017. V.Z.A., S.G.S. and
S.C.C.B. acknowledge support from FCT through Investigador FCT contracts nsº IF/00650/2015/CP1273/CT0001; IF/00028/2014/CP1215/CT0002; IF/01312/2014/CP1215/CT0004. S.G.S. also acknowledges support from FCT in the form of an exploratory project with the reference IF/00028/2014/CP1215/CT0002. J.P.F. and O.D. acknowledge support from FCT through national funds in the form of a work contract with the references DL 57/2016/CP1364/CT0005; DL 57/2016/CP1364/CT0004.

SCCB acknowledges support from  FCT through Investigador FCT contract IF/01312/2014/CP1215/CT0004.

This project has received funding from the European Research Council (ERC) under the European Union’s Horizon 2020 research and innovation programme (project {\sc Four Aces}; grant agreement No 724427).

This work has been carried out within the framework of the National Centre of Competence in Research PlanetS supported by the Swiss National Science Foundation. R.A. acknowledges the financial support of the SNSF.
This research has made use of the SIMBAD database and of the VizieR catalogue access tool operated at
CDS, France, and used the DACE platform developed in the frame of PlanetS (https://dace.unige.ch).

This work is based (in part) on data obtained via the database at the European Southern Observatory (ESO). We are grateful to all the observers of the following ESO projects, whose data we are using: 072.C-0488,  082.C-0718,  183.C-0437,  191.C-0505,096.C-0082, 099.C-0205, 099.C-0880, 65.L-0428, 66.C-0446, 267.C-5700,68.C-0415, 69.C-0722, 70.C-0044, 71.C-0498, 072.C-0495, 173.C-0606, 078.C-0829.

\end{acknowledgements}

\bibliographystyle{aa}  
\bibliography{biblio}

\onecolumn

\appendix
\appendixpage
\addappheadtotoc

\section{Including HARPS and UVES archival data: Additional figures}
\begin{figure}[h!]
        \includegraphics[width=18cm]{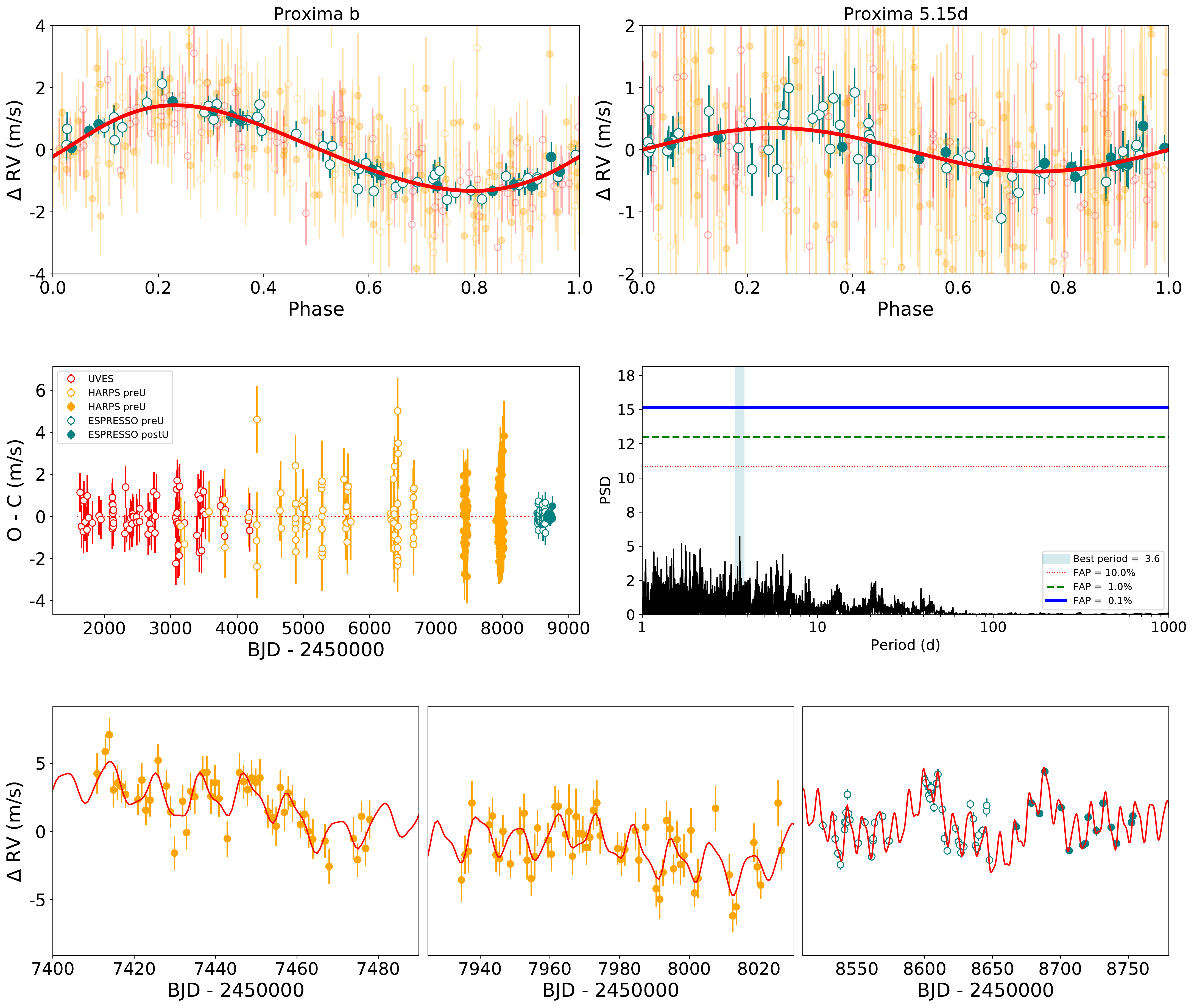}
        \caption{\textbf{Top panels:} Phase folded RV curve of the Keplerian and the 5d sinusoidal signals, after subtracting the GP component. \textbf{Middle panels: }Residuals after fitting and their periodogram.       \textbf{Bottom panel:} Zoom on the RV measurements of the different high cadence campaigns along with the best model fit from the joint GP mode of the FWHM and the RVs.}
        \label{rv_full_dataset_2p}
\end{figure}

\section{Posterior distributions}\label{ap:post_dist}

\begin{figure}[h!]
        \includegraphics[width=18cm]{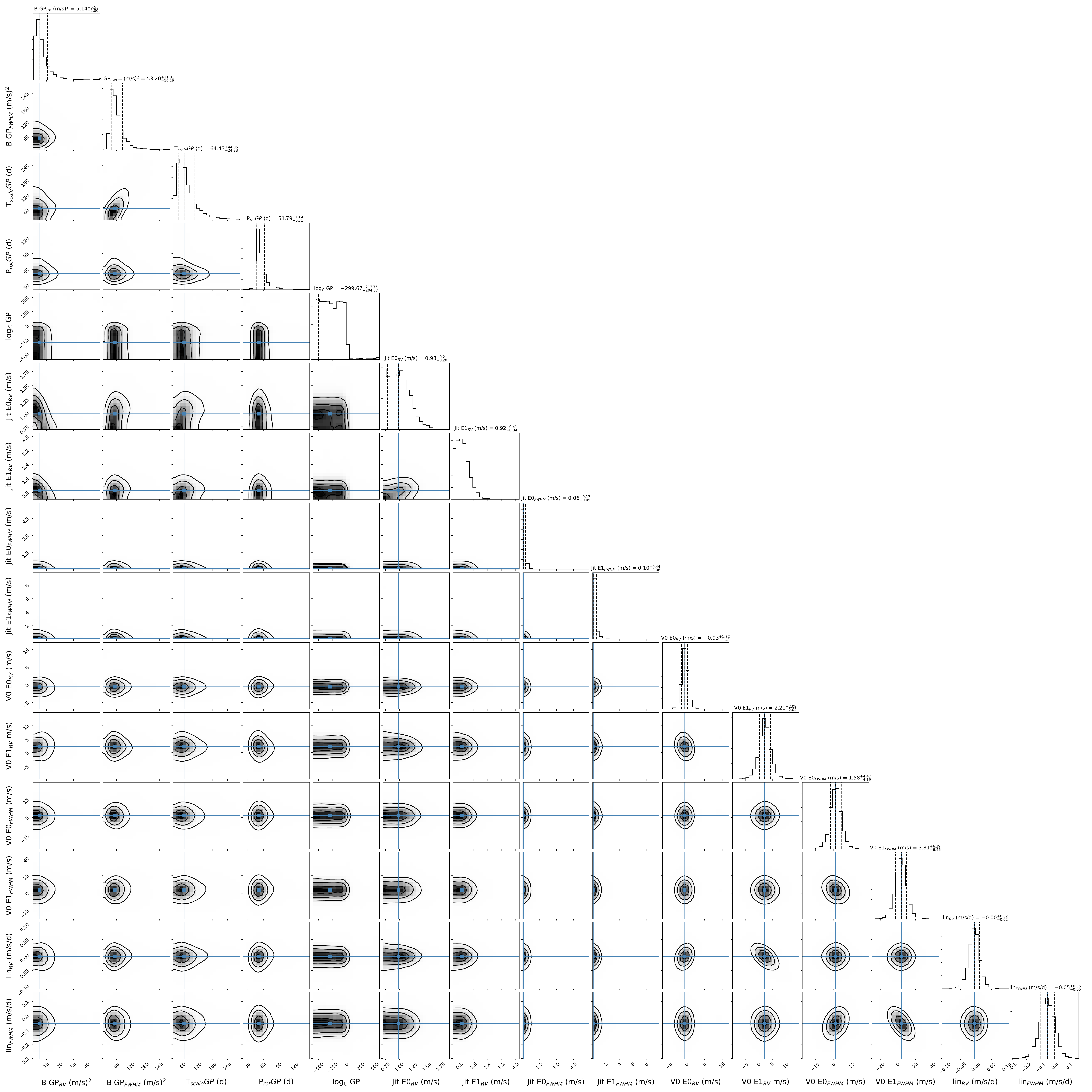}
        \caption{Posterior distributions of the parameters of the complete joint model of the FWHM and the RV data for the GP model using ESPRESSO data.}
        \label{post_dist_nop_esp}
\end{figure}

\begin{figure}[h]
        \includegraphics[width=18cm]{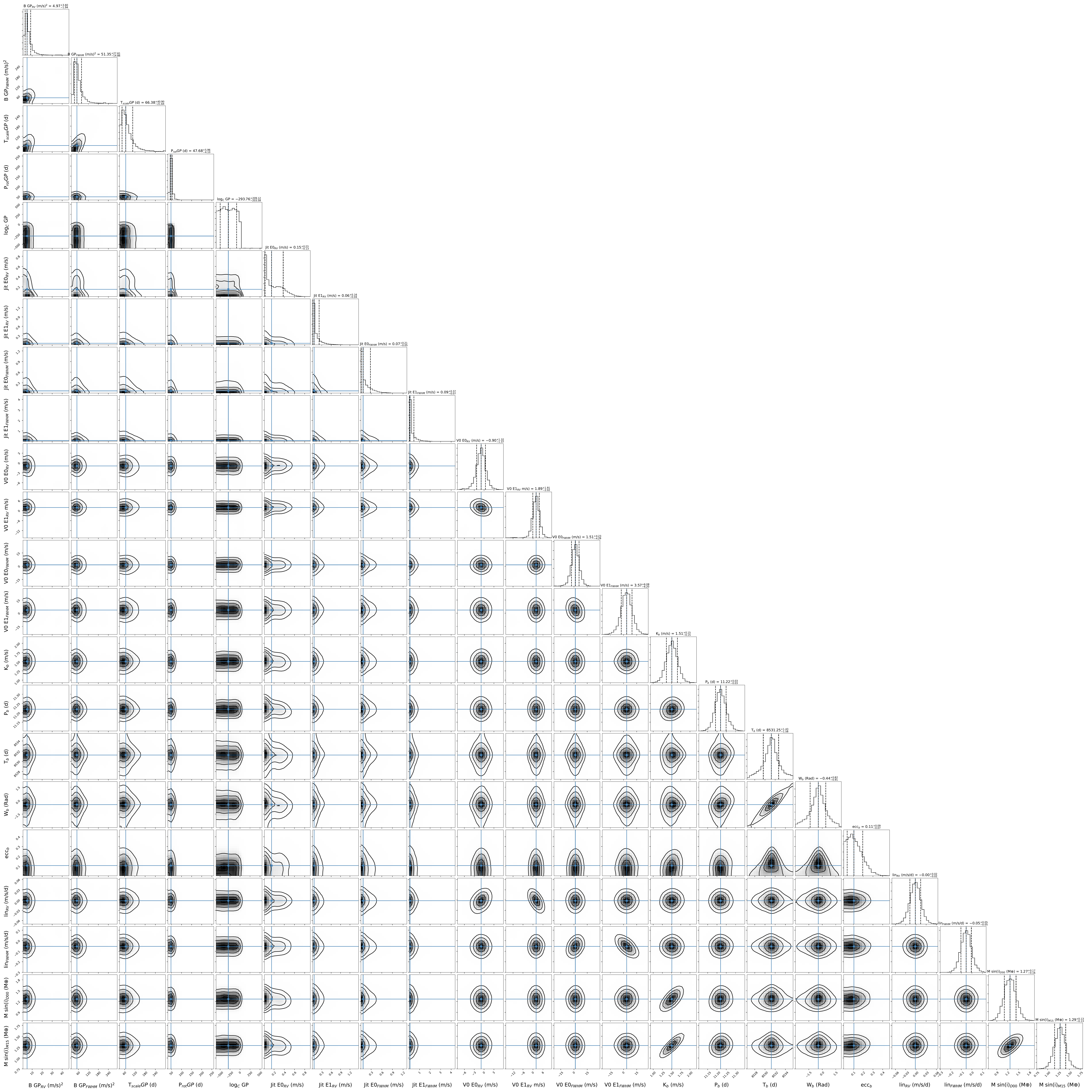}
        \caption{Posterior distributions of the parameters of the complete joint model of the FWHM and the RV data for the GP model with one extra signal using ESPRESSO data.}
        \label{post_dist_1p_esp}
\end{figure}

\begin{figure}[h]
        \includegraphics[width=18cm]{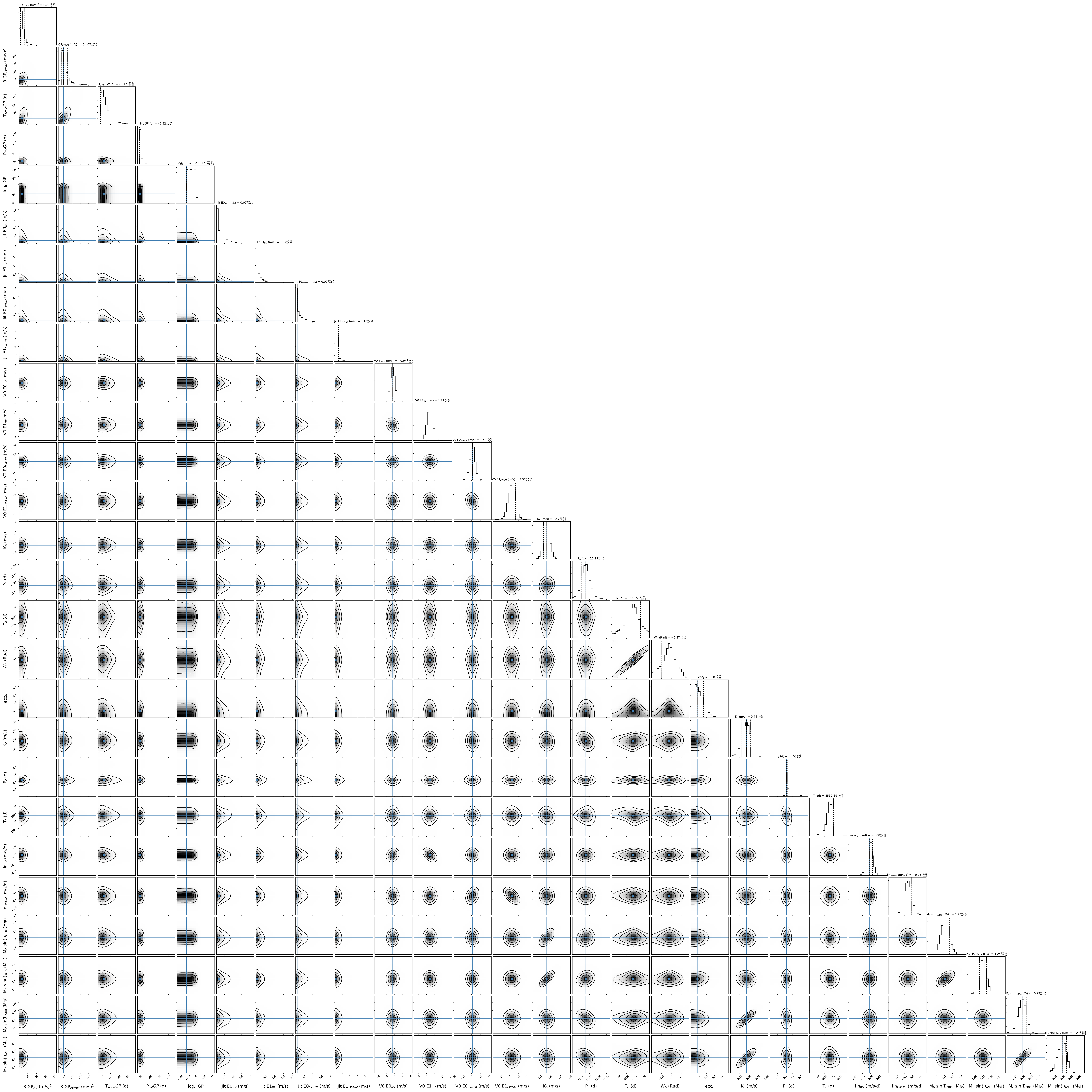}
        \caption{Posterior distributions of the parameters of the complete joint model of the FWHM and the RV data for the GP model with two extra signals using ESPRESSO data.}
        \label{post_dist_2p_esp}
\end{figure}

\begin{figure}[h]
        \includegraphics[width=18cm]{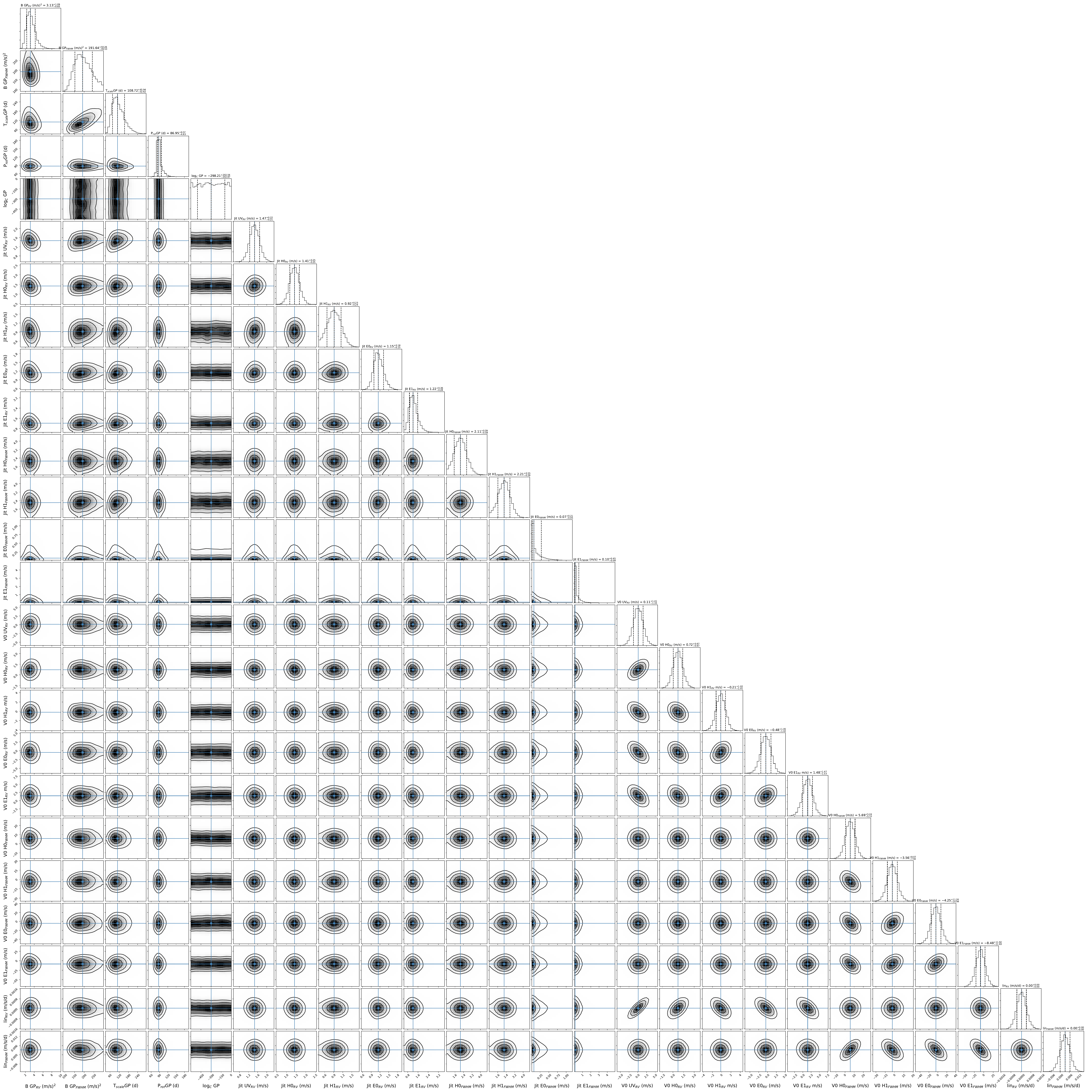}
        \caption{Posterior distributions of the parameters of the complete joint model of the FWHM and the RV data for the GP model using the full dataset.}
        \label{post_dist_nop_full}
\end{figure}

\begin{figure}[h]
        \includegraphics[width=18cm]{POST_DIST_FULL/GP_FULL_1P.pdf}
        \caption{Posterior distributions of the parameters of the complete joint model of the FWHM and the RV data for the GP model with one extra signal using the full dataset.}
        \label{post_dist_1p_full}
\end{figure}

\begin{figure}[h]
        \includegraphics[width=18cm]{POST_DIST_FULL/GP_FULL_2P.pdf}
        \caption{Posterior distributions of the parameters of the complete joint model of the FWHM and the RV data for the GP model with two extra signals using the full dataset.}
        \label{post_dist_2p_full}
\end{figure}


\end{document}